\documentclass[pdflatex,sn-basic]{sn-jnl}

\usepackage{graphicx}%
\usepackage{multirow}%
\usepackage{amsmath,amssymb,amsfonts}%
\usepackage{amsthm}%
\usepackage{mathrsfs}%
\usepackage[title]{appendix}%
\usepackage{xcolor}%
\usepackage{textcomp}%
\usepackage{manyfoot}%
\usepackage{booktabs}%
\usepackage{algorithm}%
\usepackage{algorithmicx}%
\usepackage{algpseudocode}%
\usepackage{listings}%

\begin{document}

\title[Article Title]{It Cannot Be Right If It Was Written by AI: On~Lawyers' Preferences of Documents Perceived as Authored by an LLM vs a Human}


\author*[1]{\fnm{Jakub} \sur{Harasta}}\email{jakub.harasta@law.muni.cz}

\author[1]{\fnm{Tereza} \sur{Novotná}}\email{tereza.novotna@law.muni.cz}

\author[2]{\fnm{Jaromir} \sur{Savelka}}\email{jsavelka@cs.cmu.edu}

\affil[1]{\orgdiv{Faculty of Law}, \orgname{Masaryk University}, \orgaddress{\street{Veveří 70}, \city{Brno}, \postcode{61180}, \country{Czechia}}}

\affil[2]{\orgdiv{School of Computer Science}, \orgname{Carnegie Mellon University}, \orgaddress{\street{5000 Forbes Avenue}, \city{Pittsburgh}, \postcode{15213}, \state{PA}, \country{USA}}}


\abstract{Large Language Models (LLMs) enable a future in which certain types of legal documents may be generated automatically. This has a great potential to streamline legal processes, lower the cost of legal services, and dramatically increase access to justice. While many researchers focus on proposing and evaluating LLM-based applications supporting tasks in the legal domain, there is a notable lack of investigations into how legal professionals perceive content if they believe an LLM has generated it. Yet, this is a critical point as over-reliance or unfounded scepticism may influence whether such documents bring about appropriate legal consequences. This study is the necessary analysis of the ongoing transition towards mature generative AI systems. Specifically, we examined whether the perception of legal documents' by lawyers and law students~(n=75) varies based on their assumed origin (human-crafted vs AI-generated). The participants evaluated the documents, focusing on their correctness and language quality. Our analysis revealed a clear preference for documents perceived as crafted by a human over those believed to be generated by AI. At the same time, most participants expect the future in which documents will be generated automatically. These findings could be leveraged by legal practitioners, policymakers, and legislators to implement and adopt legal document generation technology responsibly and to fuel the necessary discussions on how legal processes should be updated to reflect recent technological developments.}

\keywords{Generative AI, GenAI, Large Language Model, LLM, Automatic Text Generation, Legal Document, Perception of AI-generated Content}



\maketitle

\section{Introduction}
\label{introduction}
This paper analyzes the differences in lawyers' and law students' perceptions of documents believed to be either generated by AI or authored by humans. The aim of this study is to react to the revolution brought about by the recent advances in Large Language Models (LLMs), and provide crucial insights necessary for responsible design, development, and adoption of LLM-powered applications focused on automated generation of legal documents. Should it be known that documents are LLM-generated, potential distortions in perception present a serious concern. For example, it would be highly undesirable if a judge denies a rightful claim because they perceive the arguments in a complaint as less persuasive, knowing the document has been automatically generated. To investigate this pressing concern, we had a large group of lawyers and law students (n=75) evaluate documents labeled as AI-generated or human-crafted in terms of correctness and language quality. Further, we surveyed the participants on their beliefs on the ability of LLMs to generate legal documents automatically in the future.

ChatGPT\footnote{OpenAI: ChatGPT. Available at: https://chat.openai.com/ [Accessed 2024-08-23]} was launched on November 30, 2022, and it immediately sensitized the general public to the capabilities of LLMs to write fluent texts and lead conversations in natural language. While the foundational GPT-3 model has been available since 2020 \citep{Brown2020}, it was the accessibility of the ChatGPT service that made legal practitioners, educators, and scholars alike engage in heated discussions as to how the legal profession may change in the near future. There have been numerous examples of applications of the technology to tasks traditionally reserved exclusively for legal experts. For example, \citet{perlman2022implications} claims that ChatGPT requires re-imagining how to obtain legal services and prepare lawyers for their careers---he does so in a human-authored abstract to a scholarly article automatically generated with ChatGPT. \citet{Katz2024} tested GPT-4 on the Uniform Bar Examination (UBE), reporting the system passed the exam. \citet{Blair2024} show how even smaller LLMs can achieve near-perfect performance on basic tasks involving legal texts if they are fine-tuned. Such use cases hint at future applications of LLMs in providing legal services and increasing access to justice, namely answering legal questions, providing legal information, and, importantly, drafting legal documents.



The potential for automation, enhancement or support through LLMs is growing. Due to user-friendly access to LLMs through applications and interfaces (e.g., ChatGPT, Bard, Copilot), the experimentation with AI-based tools is available to masses desiring to become early adopters of the disruptive technology. However, the challenge of the wide use of LLMs lies in more than just their ability to achieve high (even human-like) accuracy when engaging in various tasks. If the output of LLMs is not acknowledged or perceived as accurate by the relevant actors, it may become inconsequential even if objectively accurate. The reception and perception of LLMs' output, which is tied to various societal and psychological aspects of communication, is a pressing and understudied issue.

Existing interdisciplinary research suggests generally negative sentiment towards technology, ranging from hesitancy \citep{vonEschenbach2021} to aversion \citep{Jussupow2020, Castelo2021}. Studies focusing on technology acceptance in the legal domain are scarce \citep{Xu2022, Baryse2022, Nguyen2024}. To our knowledge, no study directly focuses on examining lawyers' and law students' preferences towards documents perceived as generated by AI (specifically by ChatGPT).

To investigate if and how lawyers' and law students' perceptions of a legal document change depending on whether it is believed to be human-crafted as opposed to AI-generated, we analyzed the following research questions:
\begin{enumerate}
    \item How is the perception of a document's correctness and language quality different depending on its believed origin (human vs AI)?
    \item To what degree does experience (a law student vs a lawyer) influence the changes in perception?
    \item What are their attitudes and beliefs regarding the potential of LLMs to generate documents automatically in the future?
\end{enumerate}

Our work contributes to AI \& Law research in the following ways. To our knowledge, this is the first comprehensive study investigating differences in lawyers' perceptions of documents believed to be crafted by a human as opposed to LLM-generated. It focuses on the correctness of the documents, i.e., their potential to emulate the desired legal outcomes, and their language properties, i.e., readability or absence of grammatical/stylistic errors.

The remainder of this article is structured as follows. Section \ref{related_work} contains an overview of the related work focused on the perception of AI-generated documents and applications of LLMs in the legal domain. Section \ref{experiment} presents the experimental design, focusing on the documents, the survey used in this study, and the participating subjects. In Section \ref{results}, we present the quantitative and qualitative analysis results of the data collected from the participants. Section \ref{discussion} discusses our findings. In Section \ref{implications}, we draw implications of our research for legal practice. Section \ref{limitations} focuses on the limitations of the study. Finally, Section \ref{conclusion} summarizes the article and outlines the potential paths for future work.

To facilitate much-needed replications of our study, we release the documents and the survey used in our experiments in an accompanying online repository.\footnote{See 
\url{https://github.com/jakubharasta/AI_vs_Human_LegalDocs}}

\section{Related Work}
\label{related_work}
Experiments by \citet{Bubeck2023} and a survey by \citet{Naveed2024} have shown that LLMs offer possibilities and promises in various domains, including law. Law is based mainly on written language and is often lauded for its overwhelming production of information and documents. With the ever-increasing interest in NLP in the legal domain noted by \citet{Katz2023}, significant advances in law-related NLP research were brought by pre-trained Transformer-based Language Models (TLMs) reviewed by \citet{Greco2023}. The subsequent introduction of LLMs to the public fueled significant interest in its law-related applications, as noted by \citet{Lai2023}.

We present an overview of the related work in three broadly defined areas:
\begin{enumerate}
    \item Perception of AI-generated Content in Various Domains (Subsection \ref{topic1}) is focused on the crucial underlying issue of the perception of AI-generated content by the public and other relevant actors. The issue must be separated from the performance demonstrated by LLMs when engaging in various tasks.
    \item Automated Content Generation by LLMs in Legal Domain (Subsection \ref{topic2}) focuses on related work reporting the use of LLMs to create summaries, translations and answers to legal questions.
    \item The Use of LLMs in Legal Domain (Subsection \ref{topic3}) summarizes related work reporting the use of LLMs for various tasks, such as legal reasoning, legal research, access to justice, annotation, and legal judgment prediction.
\end{enumerate}

The related work demonstrates a need for empirical studies focusing on the perception of AI-generated documents in the legal domain. The overview below contains dozens of papers on supporting, enhancing or automating a plethora of tasks inherently perceived as requiring human attention. While the authors often report lower-than-human performance, GPT-4 significantly improves the already impressive capabilities. As a result, the use of LLMs is likely to penetrate every aspect of law. The remarkable breadth of the research conducted in the short time since November 2022 demonstrates the inevitability of AI, even in the legal domain. As such, the under-studied issue of the perception of AI-generated content in the legal domain becomes even more pressing.

\subsection{Perception of AI-generated Content in Various Domains}
\label{topic1}

Research on attitudes toward AI spans various contexts. \citet{Hancock2020} conceptualized AI-mediated communication, highlighting risks, such as undermining trust, and opportunities, such as augmenting natural communication abilities. \citet{vonEschenbach2021} observed a general hesitancy to trust AI, while \citet{Castelo2021} noted aversion toward AI echoing algorithm-related concerns raised by \citet{Jussupow2020}. These risks and opportunities are also expected to arise in the legal domain.

Increased aversion is reported when algorithms are involved in moral decision-making, particularly in domains deemed morally significant, such as medicine and law \citep{Bigman2018}. \citet{Laakasuo2021} demonstrated that decisions made by human-like robots are perceived as less moral compared to robots without resemblance to humans. Interestingly, AI-mediated communication provides flexibility in interpersonal interactions, as users are inclined to attribute part of the responsibility for negative feelings to the AI \citep{Hohenstein2020}.

Lower evaluations of AI-generated content have been reported across various domains, including Airbnb profiles \citep{Jakesch2019}, emails \citep{Liu2022}, artworks \citep{Ragot2020}, music \citep{Shank2023}, translations \citep{Asscher2023}, news articles \citep{Waddell2018} or health prevention messages \citep{Lim2024}. The level of aversion to algorithms differs between tasks perceived as subjective and objective, with perceived task objectivity leading to greater acceptance \citep{Castelo2019}. In healthcare, people tend to prefer human practitioners over AI-based technologies \citep{Miles2021}, expressing concerns over personalization of care and competence \citep{Longoni2019}. \citet{Wang2024} emphasized the necessity of incorporating essential soft skills and core principles, such as professionalism, explainability, and empathy, into GPT-based tools for medical consultations to overcome the aversion. Understanding the impact of AI-generated content on interactions within the legal domain is of utmost importance to ensure the safe deployment of AI-based tools and applications.

While the algorithms may outperform humans in specific tasks, the post-error dynamic differs significantly. \citet{Dietvorst2015} found that people lost confidence more quickly in algorithms after witnessing the same error made by both algorithmic and human forecasters. Given the inevitability of mistakes and errors, the aspect is particularly significant in the legal domain, where mistakes and errors have severe consequences for welfare or freedom.

Surveys have highlighted issues with the factuality of generated messages \citep{Wang2023}. LLMs are susceptible to hallucinations, which poses a significant challenge for further deployment in the legal \citep{Cheong2024, Magesh2024} or financial domains \citep{Kang2023Hallucination}. Information asymmetry is crucial, as those who stand to gain the most are also exposed to the most significant risks associated with hallucinations \citep{Dahl2024}. Individuals with lower literacy and education levels are at a higher risk of consuming unreliable content \citep{Oviedo2023}. \citet{Dahlkemper2023} reported that even university students with limited prior knowledge of specific issues related to their domain of expertise find evaluating the factuality of ChatGPT-generated answers challenging. These findings may manifest in the legal domain as more AI-based tools are deployed to support various legal tasks without adequately understanding the related risks.

In specific instances, the origin of the AI-generated message or advice does not have a negative effect. \citet{Logg2019} noted a preference for algorithmic advice when presented as an estimate based on the input of many individuals over human advice from a single individual. Furthermore, \citet{Leib2023} observed that advice promoting dishonesty increases dishonest behaviour, while honesty-promoting advice does not enhance honesty. The effect holds for both AI-generated and human-provided advice.

\subsection{Automated Content Generation by LLMs in Legal Domain}
\label{topic2}


\textbf{Summarization} Fine-tuning summarization models for niche domains may be prohibitively costly, which makes the use of LLMs (especially ChatGPT) appealing, particularly in the health domain \citep{Shaib2023, Tang2023}, and will most probably extend to the law as well. LLMs perform on par with human written summaries in news summarization \citep{Zhang2023, Goyal2023}.
\citet{Deroy2023, Deroy2024} explored using pre-trained abstractive summarization models and LLMs for summarization in the legal domain. They acknowledged the limitations of fully automatic approaches and proposed a human-in-the-loop approach with legal experts monitoring the quality of outcomes, addressing hallucinations and inconsistencies. LLMs have been utilized to create court decision summaries, contributing to enhanced public trust in judicial outcomes \citep{Ash2024}. AI-generated summaries offer increased accessibility and lower the cognitive effort needed for understanding, especially for non-experts. The logical progression of using LLMs includes summarizing legal and regulatory documents akin to transformer-based summarization methods \citep{Klaus2022}. \citet{Ramprasad2024} have reported efforts in this direction, focusing on the factuality of zero-shot summaries across three domains, including legal bills. Additionally, \citet{Gesnouin2024} introduced LLaMandement, a fine-tuned model for generating neutral summaries of legislative proposals. The performance of LLM-based summarization models presents a significant potential benefit to the legal community. However, the perception by its intended recipients remains largely unaddressed. 

\textbf{Translation} ChatGPT demonstrated good translation quality when powered by GPT-3.5 \citep{Karpinska2023} or GPT-4 \citep{Jiao2023}. \citet{Vieire2021} warned about the influence of machine translation in the legal domain, highlighting its under-researched impact on decision-making in critical legal situations.
Studies on the use of LLMs for legal translation are limited. \citet{Mahapatra2023} developed a parallel corpus of legislation in Indian languages and evaluated various machine translation methods, including LLMs. They noted the mediocre performance of LLMs compared to other methods. Additionally, \citet{Brivaiglesias2024} found that Google Translate outperforms LLMs, but GPT-4 is rated higher by evaluators for contextual adequacy and translation fluency. Reported performance hints at significant potential for future development and deployment. However, translating legal texts adds the language barrier to challenges associated with (lack of) specialized legal knowledge. The understudied perception of accuracy may play a significant role.

\textbf{Legal Question Answering} Given the awareness of ChatGPT among both the general and professional public, its use to generate answers to domain-specific questions is obvious. Before the introduction of ChatGPT, \citet{Metzler2021} proposed a combination of multi-task learning and zero- and few-shot learning to enhance question-answering capabilities.
The performance of LLMs in domain-specific exams attracts significant research attention. In the health domain, \citet{Kung2023} reported ChatGPT achieving or surpassing the passing threshold in three exams without fine-tuning the general model. These findings show promise in using ChatGPT in medical education and potentially clinical decision-making. \citet{Bommarito2023} analyzed LLMs' performance in an examination developed by the American Institute of Certified Public Accountants containing questions related to law, finance, accounting, technology, and ethics. The research showed that GPT-3.5 is approaching human-level performance in Remembering \& Understanding, and Application skill portions of the exam. These results led the authors to conclude that LLMs have transformative potential in the future of knowledge work. However, \citet{Shen2023} observed varying levels of ability of ChatGPT across different domains with significant underperformance in legal and science-related questions. \citet{Choi2022} investigated the use of GPT-3.5 in answering final exams in four law school courses at the University of Minnesota, noting promising but uneven performance. Overall, its question-answering performance was deemed mediocre but sufficient to earn a J.D. degree eventually. \citet{Katz2024} reported significant improvement with GPT-4, outperforming prior models and even humans in five out of seven categories of questions. Similarly, \citet{nay2024} noted improved accuracy of GPT-4 over previous models in answering multiple-choice questions about U.S. tax law. On the other hand, \citet{Martínez2024} evaluated GPT-4's UBE performance and criticized the reported values as potentially harmful in misrepresenting the actual capability of LLMs to truthfully and accurately answer legal questions. \citet{Tan2023} compared answers generated by JusticeBot, an expert tool focused on landlord-tenant disputes, and ChatGPT. They noted ChatGPT's lack of precision in providing legal information directly to laypeople. However, they suggested that combining its capabilities with expert systems may enhance question-answering performance and improve the accessibility of legal information.

\subsection{Use of LLMs in Legal Domain}
\label{topic3}


\textbf{Legal Reasoning} The survey by \citet{Huang2023Survey} notes the significant progress in NLP caused by the introduction of LLMs. However, the authors point out that the extent to which LLMs can properly reason remains unclear. 
\citet{Blair2023} evaluated GPT-3 on the statutory-reasoning dataset SARA. The authors noted apparent limitations due to imperfect prior knowledge of U.S. statutes and poor performance on synthetic statutes not encountered during training.
Studies by \citet{Nguyen2023Negation, Nguyen2023Abductive, Nguyen2023Logic} reported further limitations of using LLMs for legal reasoning. Additionally, \citet{Nguyen2023Black-Box} analysed the performance of GPT-3.5 and GPT-4 on the COLIEE Task 4 dataset. The authors raised concerns about the GPT models' ability to generalize and learn adaptable rules for unknown cases.
During the preliminary exploration of the ability of GPT-3.5 for reasoning about the FOIA requests, \citet{Baron2023} reported ChatGPT performing below the level of an experienced FOIA reviewer. On the other hand, ChatGPT exhibited the ability to bring valuable recommendations accompanied by legal reasoning.
\citet{Yu2023} suggested that the reasoning capabilities of LLMs can be significantly improved by Chain-of-Thought prompting and fine-tuning with explanations. The research indicates that the best results are achieved using prompts directly derived from specific legal reasoning techniques (e.g., IRAC).
\citet{Guha2023} prepared a robust typology for organizing legal tasks and evaluated 20 LLMs from 11 different families to provide a benchmark for the legal reasoning capabilities of LLMs. Following the modified IRAC structure, the evaluation revealed diverging performance levels, with GPT-4 emerging as the most successful model. Generally, LLMs perform better on classification tasks than those focused on application.
\citet{Kang2023IRAC} evaluated GPT-3.5's ability to conduct IRAC analysis. The research found that powerful LLMs can provide reasonable answers but mostly fail to yield correct reasoning paths.
Finally, \citet{Janatian2023} suggest using GPT-4 to extract pathways from real-world legislation to support the development of legal expert systems. Their evaluation yielded that 60\% of the generated pathways were equivalent or superior to manually created ones.

\textbf{Support for Legal Research and eDiscovery} Various tasks related to legal research, legal information retrieval, drafting or eDiscovery can be supported by LLMs.
\citet{Savelka2023Explaining} demonstrated the effectiveness of GPT-4 augmented with a legal information retrieval module, which significantly improved the accuracy of explanations of legal concepts. These findings correspond with \citet{Blair2024}, who used non-augmented models. The authors reported poor performance of most state-of-the-art LLMs (including GPT-4) in basic legal text-handling tasks, referring to them as 'sloppy paralegals'. Integrating LLMs with knowledge bases \citep{Cui2023} and other tools allowing, e.g., factual lookups \citep{Schick2023} and gathering references \citep{nakano2022webgpt} can boost performance. However, the existing research is focused on domains outside of law.
\citet{Huang2023Report} emphasized the importance of domain-specific knowledge over the general experiences distilled from ChatGPT, as demonstrated through supervised fine-tuning tasks.
\citet{Henseler2023} used GPT-4 as a supportive tool for handling data in digital investigation, entrusting it with tasks like summarization, evaluation and visualization of chat messages. Their work showcased the potential of LLMs in eDiscovery processes.
\citet{Ioannidis2023} utilized LLMs for regulatory compliance, providing identification, summarization and the impact level assessment of legal rules impacting businesses. \citet{Lam2023} employed LLMs to streamline contract drafting processes, facilitating tasks such as template selection or clause modification to suit specific contexts agreed upon by parties. Similarly, \citet{Trozze2024} used LLMs to support legal drafting in securities cases, noting models' inability to reason but leading to reduced drafting time. In the first randomized controlled trial of the effect of AI assistance, \citet{Choi2024} found slight and inconsistent improvement in the quality of legal analysis but significant and consistent increases in speed.

\textbf{Access to Justice/Law} Pre-trained language models offer a significant potential for increasing the accessibility of domain-specific knowledge to laypeople. Classical information retrieval can be paired with pre-trained models to improve domain expert advice substantially. \citet{Metzler2021} introduced a consolidated model combining multi-task learning and zero- or few-shot learning, which may serve as a foundation for including ChatGPT into various pipelines for improved question-answering capabilities. Addressing the procedural and financial barriers to legal services, LLMs can potentially improve access to justice and government services \citep{Bommasani2022}.
Research has explored various avenues for utilizing LLMs, including support to lobbying activities \citep{Nay2023}, mediation support \citep{Westermann2023Mediator}, and mapping laypeople's fact-based narrative to legal issues \citep{Westermann2023Layperson}. \citet{Chien2024} highlighted the deployment of GPT-powered chatbots in the judiciary, improving efficiency in addressing the legal needs of low-income groups.
\citet{Goodson2023} reported a proof-of-concept utilization of LLMs to streamline the legal aid intake process and support the initial legal triage. Such use addresses a known bottleneck in legal aid delivery.
Laypeople often utilize LLM-based tools for self-help in legal issues. They mainly seek statements about applicable laws, procedural guidance and directories of potential services or webpages delivering further assistance \citep{Hagan2024}.

\textbf{Legal Judgment Prediction} Legal judgment prediction aims to predict judgment results automatically. Experiments showcase the potential of LLMs for this task. 
\citet{Hamilton2023} demonstrated GPT-2's ability to achieve better-than-random accuracy in simulating judicial rulings of the U.S. Supreme Court from 2010 to 2016. Similarly, zero-shot prediction on the Chinese criminal case dataset using GPT-3 reported by \citet{Jiang2023} outperformed the baseline.
The investigation by \citet{Trautmann2022} focused on the prediction of judgments of the European Court of Human Rights and the Federal Supreme Court of Switzerland. The study found that the LLMs exhibited superior zero-shot predictive capabilities compared to baseline models. However, the reported results fell short of existing supervised approaches. \citet{Wu2023} demonstrated the possibility for improved performance by combining LLMs with domain models.
The introduction of GPT-4 presented a significant development, yielding remarkably superior prediction results compared to previous models and existing approaches \citep{Shui2023}.

\textbf{Annotation} LLMs can automatically annotate textual elements, significantly reducing the effort and costs associated with empirical analysis and legal review. \citet{Drapal2023} proposed using GPT-4 to support thematic analysis in empirical legal studies with promising results in zero-shot classification and the potential for further improvement with expert feedback.
\citet{Savelka2023Unlocking} and \citet{Savelka2023Effectiveness} evaluated GPT-3.5's effectiveness in contract review, statutory and regulatory provisions investigation, and case-law analysis. \citet{Gray2024} utilized GPT-4 and GPT-4-Turbo to create sentence-level annotations for factor analysis. \citet{Savelka2023Textual} found GPT-4 comparable to well-trained student annotators in analyzing textual data.
\citet{Oliveira2024} explored the use of GPT-3 in a pipeline integrating LLMs and weak supervision for named entity recognition tasks. They observed comparable performance to traditional manual text labelling methods at a lower cost.

\section{Evaluation Experiment}
\label{experiment}

We conducted an evaluation experiment of two human-crafted documents (Brief and Verbose) detailed in Subsection \ref{documents} and two groups of evaluators described in Subsection \ref{participants}. Group A received the Brief document labeled as AI-generated and the Verbose one labeled human-crafted. Group B received the documents with opposite labelling. Participants were instructed to evaluate the language quality and correctness of the documents through a survey. They were allowed to provide open-ended comments to their evaluation. Details about the survey can be found in Subsection \ref{survey}.

\subsection{Documents}
\label{documents}
We prepared two written acknowledgements of debt. Both documents were human-crafted without the use of LLMs and drafted to fit on one page. One document was prepared to be as brief as possible (Brief), while the other was drafted to contain some additional information (Verbose). Both documents were prepared to comply with standard practices (e.g., designation of parties, legal jargon used). 

The \textbf{Brief document} contained the headline 'Acknowledgement of Debt' followed by the designation of both the debtor and the creditor (name, surname, date of birth, address). The following part outlined the origin of the debt (failure to pay the lease). Subsequently, the document contained an explicit acknowledgement of debt required by law and the due date. Finally, there was a confirmation that the acknowledgement was not written under duress. The Brief document was approximately 200 words in length.

The \textbf{Verbose document} contained the structured designation of the debtor and the creditor (name, surname, date of birth, address) followed by the headline 'Acknowledgement of Debt'. The document contained an explicit acknowledgement of debt required by law. Subsequently, the document outlined the origin of the debt (failure to pay the lease) in a structured and more detailed manner. Finally, the document contained the due date and the confirmation that it was not written under duress. The Verbose document was approximately 250 words in length.

\begin{figure}
\centering
\includegraphics[width=0.495\textwidth]{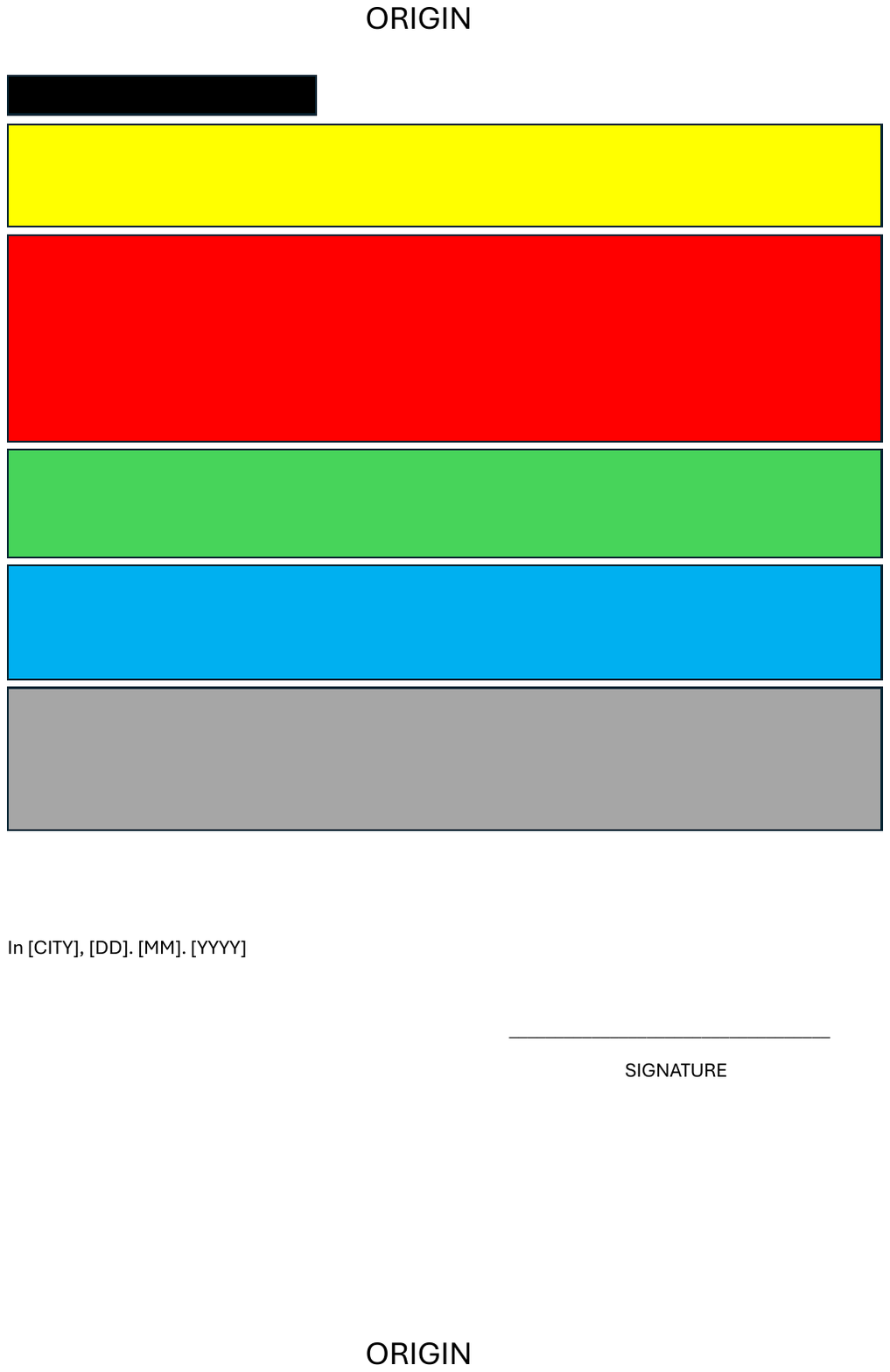}
\includegraphics[width=0.495\textwidth]{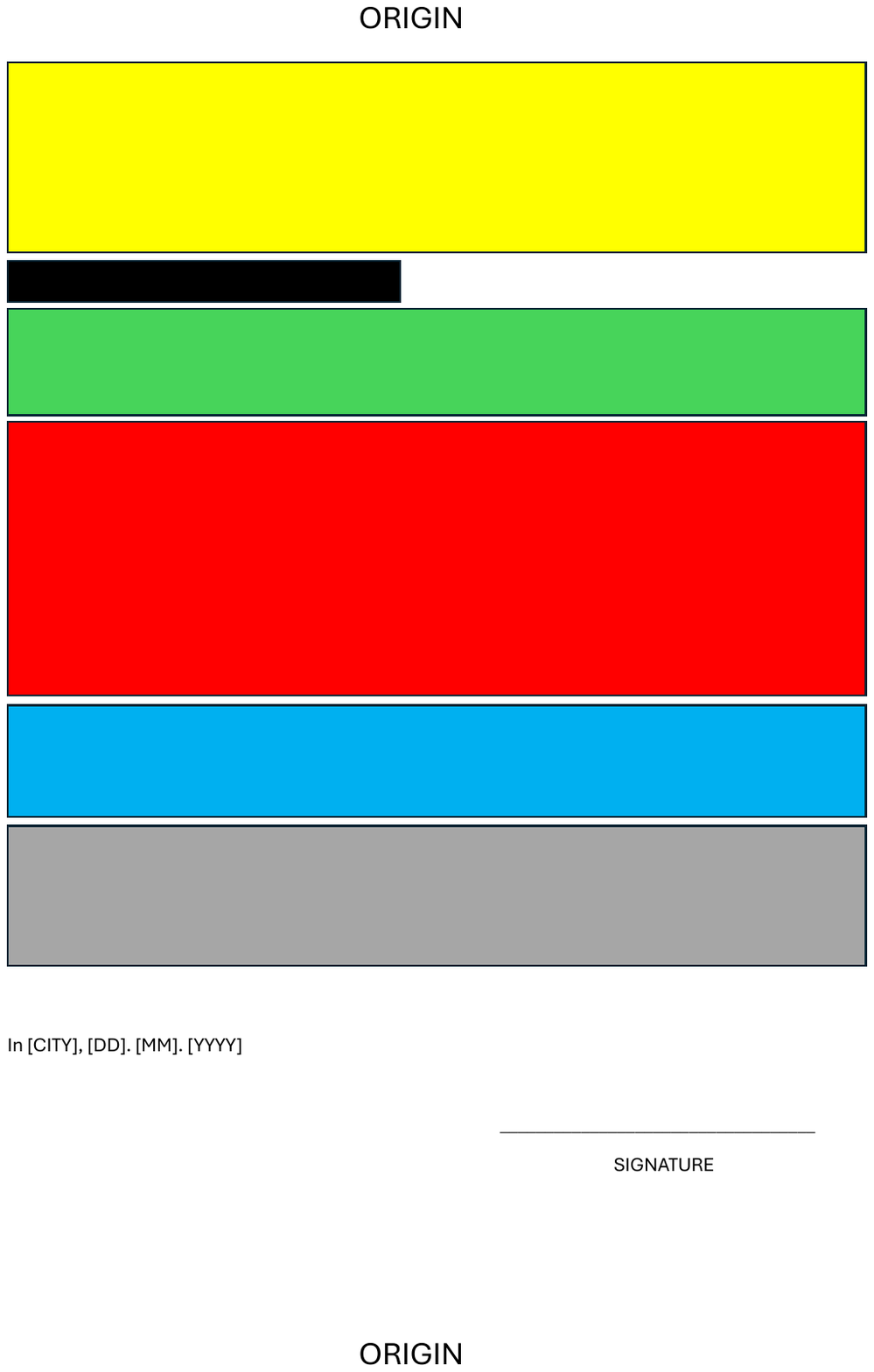}
\caption{The figure outlines the structure of the Brief (left) and the Verbose (right) documents. Structure contains \colorbox{yellow}{designation of parties}, \colorbox{black}{\textcolor{white}{headline}}, \colorbox{green}{acknowledgement of debt}, \colorbox{red}{origin of debt}, \colorbox{cyan}{\textcolor{white}{due date}} and \colorbox{lightgray}{confirmation of absence of duress}.}
\label{doc-structure}
\end{figure}

Further, each document was modified into two versions. One version was labeled 'AI-GENERATED DOCUMENT' in both the header and the footer in blue highlight. The other version was labeled 'HUMAN-CRAFTED DOCUMENT' in both the header and the footer in yellow highlight. We avoided using green and red highlights for their well-known associations to correct and incorrect options.

\begin{figure}
\centering
\includegraphics[width=0.495\textwidth]{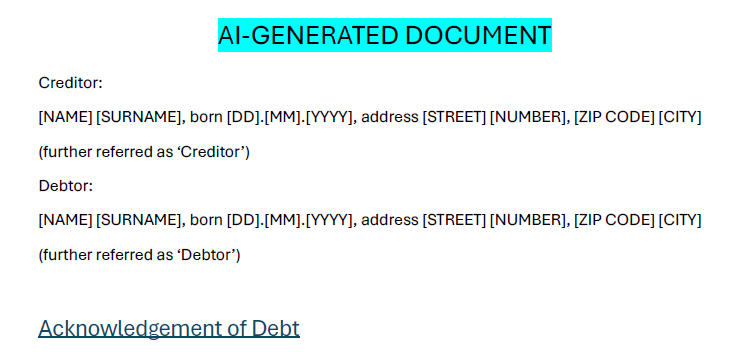}
\includegraphics[width=0.495\textwidth]{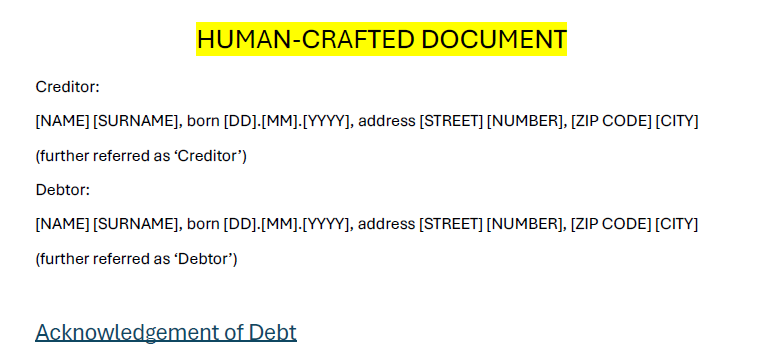}
\caption{The figure contains snippets of two variants of the Verbose document. One is designated as 'AI-GENERATED DOCUMENT' (left) in its header, and the other as 'HUMAN-CRAFTED DOCUMENT' (right). The same designation appears also in the footer of every variant.}
\label{doc-origin}
\end{figure}

The structure of both documents and the relative length of their parts are outlined in Figure \ref{doc-structure}. The designation distinguishing between two variants of the same document is shown in Figure \ref{doc-origin}.

\subsection{Participants}
\label{participants}
We recruited participants through a mailing campaign at Masaryk University in Czechia and calls for participants distributed via Facebook and X, reaching approximately 1,200 people.

Prior to taking the survey, prospective participants were informed about the scope of their participation, data usage, and rights, including the right to withdraw at any time. The data was collected anonymously, and no personally identifiable information was shared with the research team. No piece of collected data can be tied back to a specific individual.

A total of 89 prospective participants expressed their interest in participating. Of the 89 participants, 53 were law students enrolled in a law degree-granting program, and 36 were lawyers. We used stratified random sampling to divide participants into Group A and Group B.

Group A comprised 26 law students and 18 lawyers, totaling 44 participants. We received completed surveys from 39 participants (88.6\%) -- 22 law students (84.6\%) and 17 lawyers (94.4\%). Five participants failed to complete the survey.

Group B comprised 27 law students and 18 lawyers, totalling 45 participants. We received completed surveys from 36 participants (80\%) -- 20 law students (74.1\%) and 16 lawyers (88.9\%). Nine participants failed to complete the survey.

Initially, we planned to discard the answers from participants who completed the survey in under ten minutes. However, we did not discard any answers for this reason, as all of the participants took longer to finish the survey.

\subsection{Survey}
\label{survey}
The survey was conducted online via Microsoft Forms. Group A was presented with the Brief document labeled as AI-generated and the Verbose document labeled as human-crafted. Group B was presented with the Brief document labeled as human-crafted and the Verbose document labeled as AI-generated. Participants were unaware that both documents were human-crafted and that no AI-based tools were used. The survey was distributed to all the participants on January 30, 2024. Participants were requested to finish their evaluation by February 11, 2024.

Participants were to score the documents on a scale of 1 (worst) to 5 (best) in terms of the following categories:
\begin{enumerate}
    \item \textit{Language Quality}: The degree to which the document conforms with the language expectations associated with legally binding documents. The participants were tasked to check for grammatical and stylistic errors or the use of inappropriate words within a given context.
    \item \textit{Correctness}: The degree to which the document is correct. The participants were to evaluate the fulfilment of legally required conditions and factual and formal coherence. 
\end{enumerate}

Participants were also asked to provide brief explanations (suggested length of up to 100 words) for each score given. Finally, the evaluation included an open-ended question to collect participants' opinions on the feasibility of generating debt acknowledgements automatically and achieving output quality comparable to that of humans.

In summary, the participants encountered the following questions:
\begin{itemize}
    \item Question 1: Score the language quality of the AI-generated document
    \item Question 2: Briefly explain the score given in Question 1
    \item Question 3: Score the correctness of the AI-generated document
    \item Question 4: Briefly explain the score given in Question 3
    \item Question 5: Score the language quality of the human-crafted document
    \item Question 6: Briefly explain the score given in Question 5
    \item Question 7: Score the correctness of the human-crafted document
    \item Question 8: Briefly explain the score given in Question 7
    \item Question 9: After comparing the AI-generated and human-crafted documents, assess to what extent full automation is possible and the human-level performance achievable
\end{itemize}

\noindent Before the survey could be submitted, participants were required to answer all four scoring questions (Questions 1, 3, 5, and 7). Answering open-ended questions (Questions 2, 4, 6, 8, and 9) was not mandatory.

We evaluated the survey using both quantitative and qualitative methods. Quantitatively, we analyzed participants' scores, focusing on average scores and preferences for documents labeled as human-crafted or AI-generated, with a further distinction between the Brief and the Verbose documents. Qualitatively, we examined open-ended responses.

The responses to the open-ended questions were subjected to a qualitative thematic analysis methodologically following \citet{Braun_Clarke_2006} and applied, for instance, in \citet{Liffiton_Sheese_Savelka_Denny_2023}. The thematic analysis was conducted by grouping responses to open-ended questions according to the characteristics assessed - language and correctness - across documents (human-crafted, AI-generated) and groups (A and B). As a result, two sets of responses were created: one consisting of responses to the language quality and the other to responses to the correctness. These two sets of responses were thematically coded, i.e. we were looking for similar elements across the set of responses. These elements were then assigned a sentiment, i.e., whether they were mentioned positively or negatively. The elements were then grouped into metacategories, called themes, according to their situational and semantic proximity. Thus, a schema of themes emerged, defined by their sub-elements. In addition, each theme can have a positive or negative attribute.

\section{Results}
\label{results}

\subsection{Document Preference}
\begin{figure}
\centering
\includegraphics[width=0.495\textwidth]{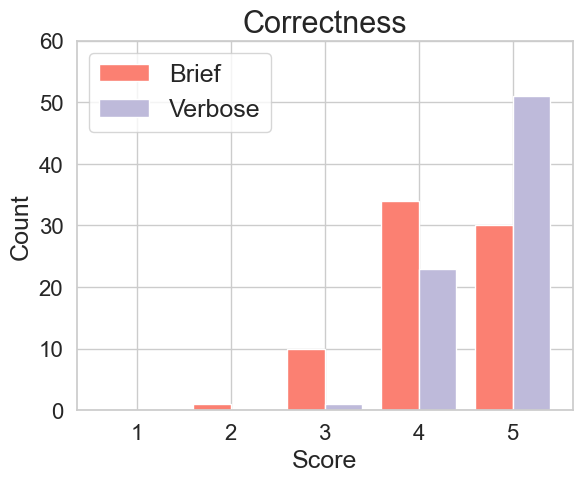}
\includegraphics[width=0.495\textwidth]{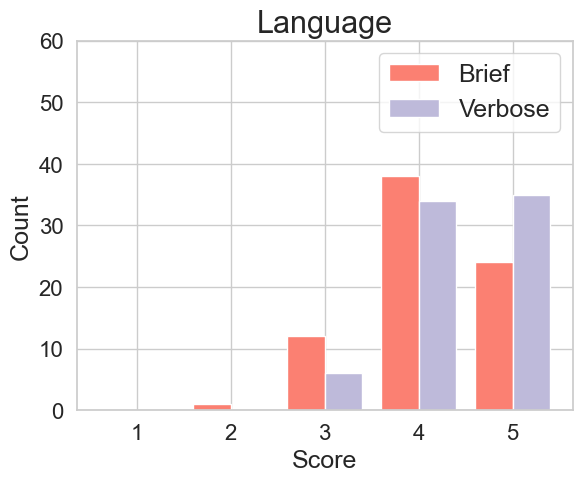}
\includegraphics[width=\textwidth]{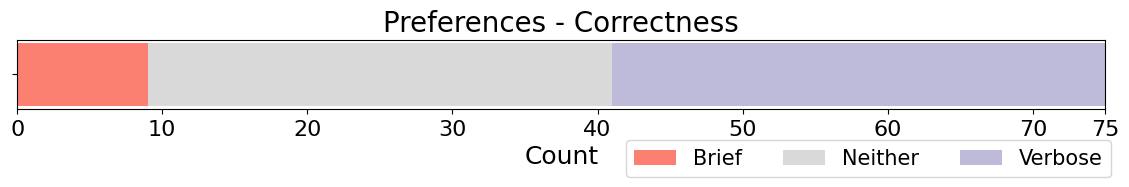}
\includegraphics[width=\textwidth]{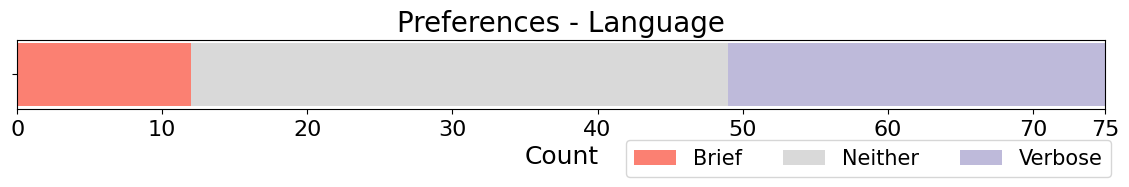}
\caption{The figure summarizes the participants' preferences between the two documents (Brief and Verbose). The top two charts show the distribution of scores awarded to each document in terms of their Correctness and Language Quality (1--worst; 5--best). The bottom two charts present the results of the side-by-side comparisons, showing how many times each of the documents was preferred (if any). Overall, a clear preference for the Verbose document over the Brief one can be observed.}\label{document-type}
\end{figure}

While both documents were authored to be of comparable quality, it appears that the participants preferred the Verbose one, especially in terms of correctness. The mean overall correctness rating of the Brief document was 4.24 (39$\times$ presented as AI-generated and 36$\times$ as human-crafted) compared to the 4.67 of the Verbose one (36$\times$ AI and 39$\times$ human). Seen side-by-side, the participants judged the Verbose document as more correct than the Brief one 34 times (out of 75). The Brief document was deemed as more correct only 9 times. In the remaining 32 times the documents were evaluated as comparable. In terms of language quality, the trend was still there (4.39 Verbose versus 4.13 Brief), where the Verbose document was preferred 26 times compared to 12 preferences for the Brief one (37 neither). Figure \ref{document-type} provides additional detail into how the participants rated the two documents. While the comparison of the documents is not the subject of this study, it presents an important aspect that needs to be considered when interpreting the results below.

\subsection{Correctness}
\label{correctness}
\begin{figure}
\centering
\includegraphics[width=0.495\textwidth]{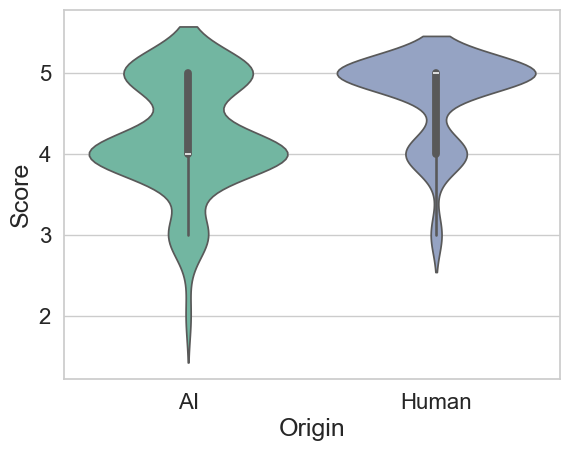}
\includegraphics[width=0.495\textwidth]{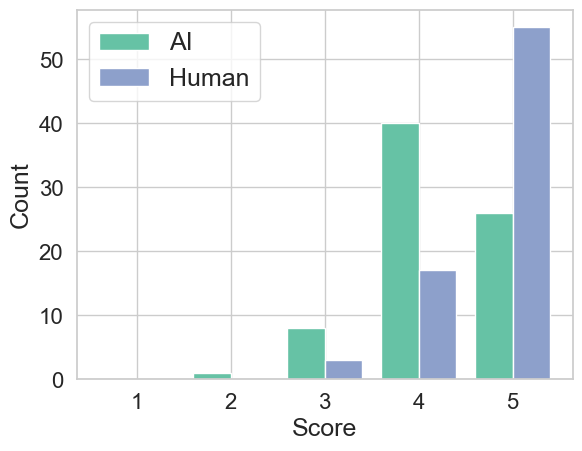}
\includegraphics[width=\textwidth]{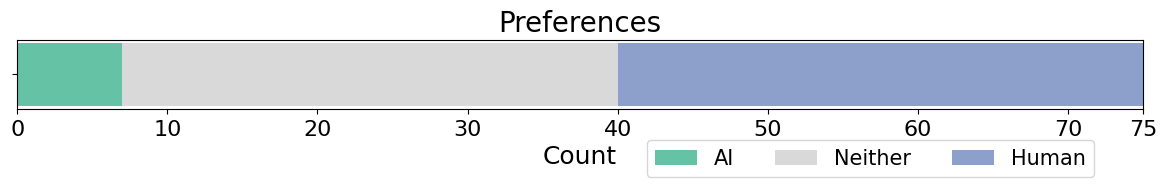}
\caption{The figure summarizes the participants' preferences between the documents when labeled as AI-generated versus human-crafted in terms of their Correctness. The top two charts show the distribution of scores awarded to the documents carrying the ``AI`` (green) or ``human`` (blue) labels (1--worst; 5--best). The bottom chart presents the results of the side-by-side comparison, showing how many times each of the labels was preferred (if any). Overall, a clear preference for the documents labeled as human-crafted over those labeled as AI-generated can be observed.}\label{corr-overall-dist}
\end{figure}

Figure \ref{corr-overall-dist} shows the overall evaluation of the documents in terms of correctness, focusing on whether a document was labeled as human-crafted or AI-generated. The participants preferred a document when it was presented as human-crafted over a document labeled as AI-generated. Specifically, the mean evaluation of the two documents when labeled as human-crafted was 4.69, compared to 4.21 when the same documents were marked as AI-generated. The side-by-side comparison presents a clear message when the human-crafted designation yielded 35 preferences over the mere 7 of the documents perceived as generated by AI (33 neither). This difference is statistically significant by Fisher exact test ($p<10^{-4}$ using Bonferroni correction for 4 comparisons).

\begin{figure}
\centering
\includegraphics[width=0.495\textwidth]{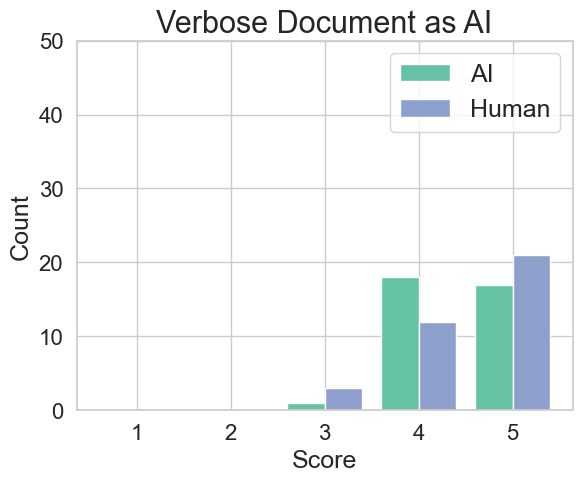}
\includegraphics[width=0.495\textwidth]{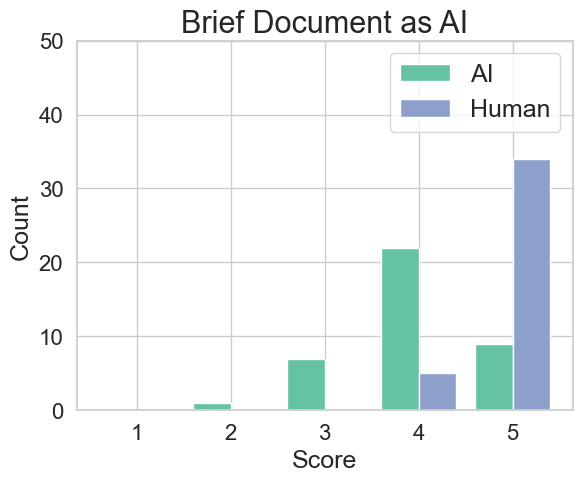}
\includegraphics[width=\textwidth]{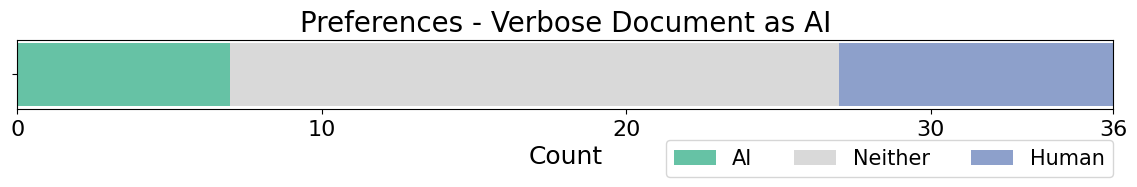}
\includegraphics[width=\textwidth]{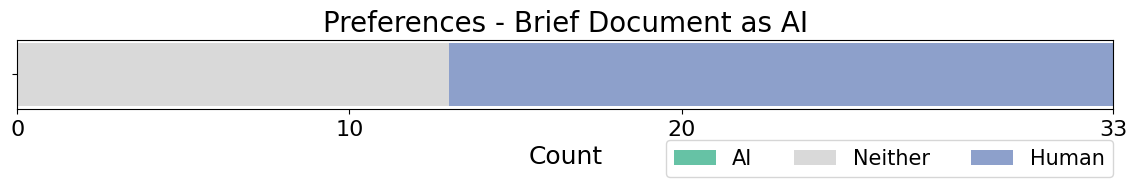}
\caption{The figure summarizes the participants' preferences between the documents when labeled as AI-generated versus human-crafted in terms of their Correctness when the document is taken into account (Brief and Verbose). The top two charts show the distribution of scores awarded to the documents carrying the ``AI`` (green) or ``human`` (blue) labels (1--worst; 5--best). The bottom charts present the results of the side-by-side comparisons, showing how many times each of the labels was preferred (if any) per document. While the Verbose document is clearly preferred overall, the AI-generated label appears to largely mitigate the effect in case of when attached to the Verbose document and largely amplify it when put on the Brief document.}\label{corr-document-dist}
\end{figure}

Figure \ref{corr-document-dist} provides additional insight into the preferences, focusing on the interaction between the AI-authored and human-crafted designation and the individual documents. While the Verbose document appears to be preferred by the participants overall (Figure \ref{document-type}), the effect is negated if it is marked as AI-generated and the Brief document as authored by a human. In that case, we can even observe that the Brief document was deemed slightly more correct on average (4.5 Brief versus 4.44 Verbose). In terms of the side-by-side comparisons, the Verbose document (labeled as AI-generated) was preferred 7 times, whereas the one marked as authored by a human was deemed more correct 9 times (20 neither). The overall preference for the Verbose document was significantly boosted when this one was labeled as produced by a human. Specifically, the mean correctness rating of the Verbose document marked as coming from a human was 4.87 compared to the mere 4.0 of the Brief document presented as AI-generated. The Verbose document was preferred in 26 out of 39 instances, while the Brief document was not preferred a single time (i.e., 13 neither was preferred).

\begin{figure}
\centering
\includegraphics[width=0.495\textwidth]{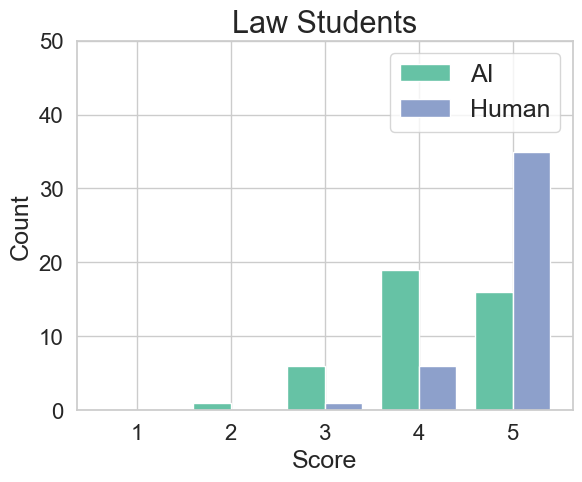}
\includegraphics[width=0.495\textwidth]{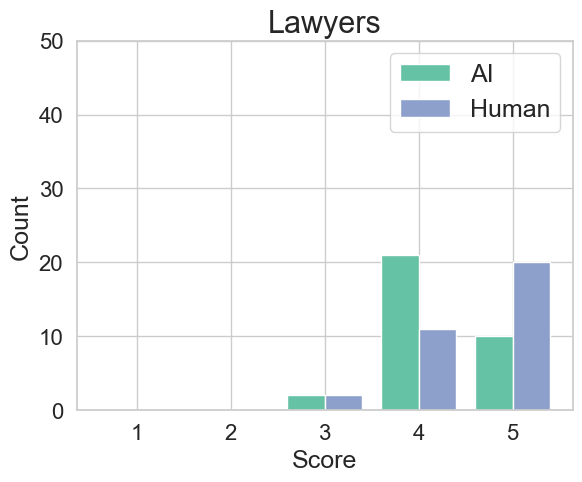}
\includegraphics[width=\textwidth]{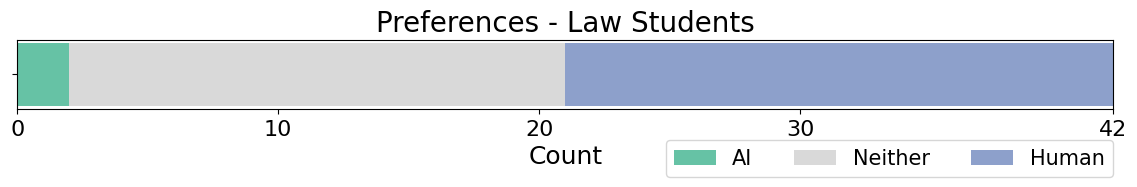}
\includegraphics[width=\textwidth]{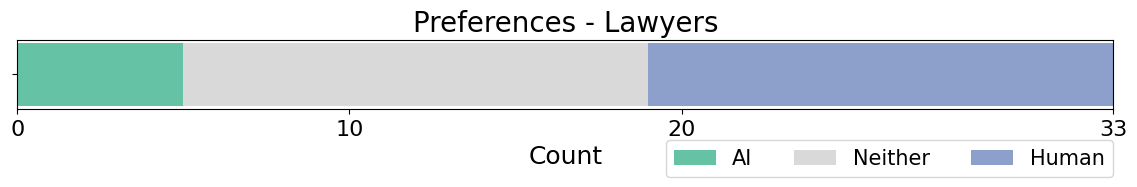}
\caption{The figure summarizes the participants' preferences between the documents when labeled as AI-generated versus human-crafted in terms of their Correctness when the participants' background is taken into account (a law student and a lawyer). The top two charts show the distribution of scores awarded to the documents carrying the ``AI`` (green) or ``human`` (blue) labels (1--worst; 5--best). The bottom charts present the results of the side-by-side comparisons, showing how many times each of the labels was preferred (if any) per participants' background. It appears the AI-generated label has much larger effect on the law students as compared to the lawyers.}\label{corr-background-dist}
\end{figure}

Figure \ref{corr-background-dist} focuses on the interaction between the participants' background (i.e., law student versus lawyer) and a document's AI/human designation. Note that the result is not statistically significant based on the Fisher exact test ($p=0.2221$). However, it appears that the designation of origin somewhat more influenced law students, preferring a document marked as human-crafted (4.81 labeled as human versus 4.19 for AI). Specifically, the law students preferred the document marked as coming from a human in 21 instances compared to only 2 preferences for a document presented as generated by AI (19 neither). Lawyers appear to have been less susceptible to the AI/human designation while still preferring a document when presented as coming from a human (4.55 human vs 4.24 AI). They preferred a human-crafted document in 14 cases compared to 5 instances when a document marked as AI-generated was preferred (14 neither).

To further understand the participants' preferences, we conducted a thematic analysis of their explanations of the submitted document scores according to the methodology described in Subsection \ref{survey}.  We identified five prominent themes in the explanations of scores regarding the correctness of the documents:

\begin{itemize}
    \item \emph{Legal requirements} -- whether the document meets legal standards and requirements.
    \item \emph{Coherence and logical structure} -- references to relevant parts of the document, accuracy of internal references, formal structure of the document,  redundancy.
    \item \emph{Formal correctness} -- accurate dating, establishment of terms and conditions, identification of parties and contracts (external references).
    \item \emph{Expected legal consequences} -- whether it is clear from the document what legal consequences may be drawn from it.
    \item \emph{Wording and conventions} -- whether the document follows common legal conventions and terms.
\end{itemize}

\noindent All the themes were either mentioned in a positive or a negative sense. Examples of positive mentions are e.g.,
\begin{quote}
\begin{itemize}
    \item[] ``Fulfils legal requirements.''
    \item[] ``Correctly identifies the parties, the related documents (contract) and the correct Act and its Section.''
\end{itemize}
\end{quote}

Examples of negative mentions are e.g.,

\begin{quote}
\begin{itemize}
    \item[] ``Document does not contain the specific provisions of the referenced contract.''
    \item[] ``Paragraphs appear inconsistent, and information is not conveyed smoothly.''
\end{itemize}
\end{quote}

Upon closer examination of the individual occurrences of these thematic categories, we see the following. Compliance with \textit{legal requirements} is mentioned exclusively positively. This is because both documents objectively and deliberately meet the legal requirements. This topic is often mentioned in the context of the relevant Section 2053 of the Czech Civil Code, which sets out the conditions for the acknowledgement of the debt, e.g.,
\begin{quote}
\begin{itemize}
    \item[] "The document contains what it should contain according to the relevant provision [of the Civil Code]."
    \item[] "It fulfils the conditions in Section 2053 of the Civil Code".
\end{itemize}
\end{quote}

Furthermore, the theme is often mentioned in the context of \textit{expected legal consequences}, i.e.,
\begin{quote}
\begin{itemize}
    \item[] "The document meets the requirements of Section 2053, and the document should trigger the expected legal consequences of debt acknowledgement."
    \item[] "The document meets the requirements of debt acknowledgement, i.e. it contains (i) the reason for the debt (debt from the lease agreement) and (ii) the amount owed (CZK 25,000). The document contains placeholders for identifying the debtor and the creditor, the completion of which will produce the desired legal consequences once the document is signed."
\end{itemize}
\end{quote}

Similarly, the theme also appears frequently in combination with the theme of \textit{formal correctness} in a positive sense. The responses showed that there is a broad consensus that formal correctness is linked to expected legal consequences:
\begin{quote}
\begin{itemize}
    \item[] "The debt is acknowledged by the document to the extent and in the amount required by Section 2053 of the Civil Code, and this is done with sufficient precision."
    \item[] "I consider the requirements of Section 2053 of the Civil Code to have been fulfilled. The document is materially and formally coherent."
\end{itemize}
\end{quote}

The theme of \textit{coherence and logical structure} contains negative comments on redundancy or insufficient clarity of text, e.g.
\begin{quote}
\begin{itemize}
    \item[] "[...] the correct terminology is used, although some of the wording seems redundant."
    \item[] "I had to think about which rent for which month is due."
    \item[] "In the document's introduction, we learn that the debt was incurred as rent due on 15th January 2015, and then we learn that the debtor did not pay the rent until 2023; as a result, the document  appears less consistent."
\end{itemize}
\end{quote}

In a positive sense, the document is commented on as written clearly and well, proceeding logically and clearly and emphasizing the important facts.

\textit{Formal correctness} is the most frequently mentioned theme. It includes correct dates and times, terms and conditions, identification of parties and related documents (contracts) or other external references. \textit{Formal correctness} also contains comments on the completeness or incompleteness of the information included, not only those required by the legal regulation - thus, compared to the theme of \textit{legal requirements}, \textit{formal correctness} is more general and broader. In a positive sense, e.g.:

\begin{quote}
\begin{itemize}
     \item[] "The document contains all relevant information (although it does not explicitly state that the debtor is the tenant, but the content makes this clear)."
     \item[] "It seems to me that the document correctly and consistently identifies the relevant statutory provision, the parties to the contract, and sufficiently defines the reason and the amount of the acknowledged debt."
\end{itemize}
\end{quote}

On the other hand, negative comments appeared to focus on a lack of detail, e.g.,

\begin{quote}
\begin{itemize}
     \item[] "The document does not identify the creditor and the debtor concerning the lease agreement."
     \item[] "It does not refer directly to the article of the agreement, where the obligation to pay rent is contained."
\end{itemize}
\end{quote}

\textit{Formal correctness} was also assessed regarding inaccuracies or inconsistencies. Positive mentions included a lack of any inaccuracies and inconsistencies. On the other hand, some participants reported negatively emphasizing inconsistencies in referring to parties and legal provisions or lacking details about the debt payment method.

As mentioned above, the \textit{expected legal consequences} are often mentioned in the context of compliance with \textit{legal requirements}, as in practice, these two situations are logically linked. As with \textit{legal requirements}, the theme is only mentioned positively because if documents objectively meet the legal requirements, there must be expected legal consequences. The theme includes comments on whether the legal consequences are apparent from the document and the possible use of the documents in practice. Comments such as

\begin{quote}
\begin{itemize}
     \item[] "[...] and it is also obvious what legal effects it intends to cause."
     \item[] "It could probably be used in practice after minor modifications."
     \item[] "[It is] applicable in practice."
\end{itemize}
\end{quote}

The final topic concerning the correctness of the document is that of \textit{wording and conventions}. This is defined simply as the document following the usual conventions in wording and use of legal language (in general). The theme is mentioned mostly negatively, with a few positive exceptions. It is only discussed positively when a Verbose document is labeled as human-crafted. In that case, it is described as 

\begin{quote}
\begin{itemize}
     \item[] "More expertly worded than an automated document."
     \item[] "The wording is concise, spare, but consistent in a way that leaves no room for doubt."
\end{itemize}
\end{quote}

In a negative sense, comments focused on inappropriate form, lack of professionalism, or poor choice of words. Examples include, 

\begin{quote}
\begin{itemize}
     \item[] "The form is skeletal."
     \item[] "The sentence 'This is a debt of 25 000 CZK' does not appear professional."
     \item[] "The sentences are oddly worded to be absolutely certain about the document's meaning."
     \item[] "The last paragraph appears 'artificial' but contains identical information to a human-written document."
\end{itemize}
\end{quote}

\begin{figure}
\centering
\includegraphics[width=0.495\textwidth]{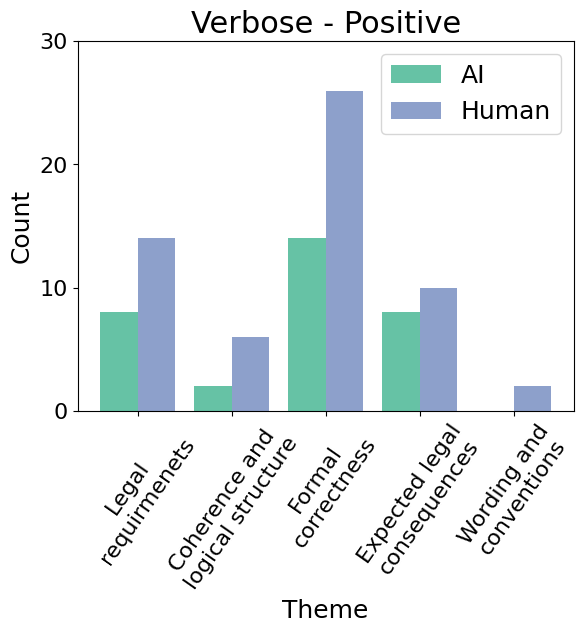}
\includegraphics[width=0.495\textwidth]{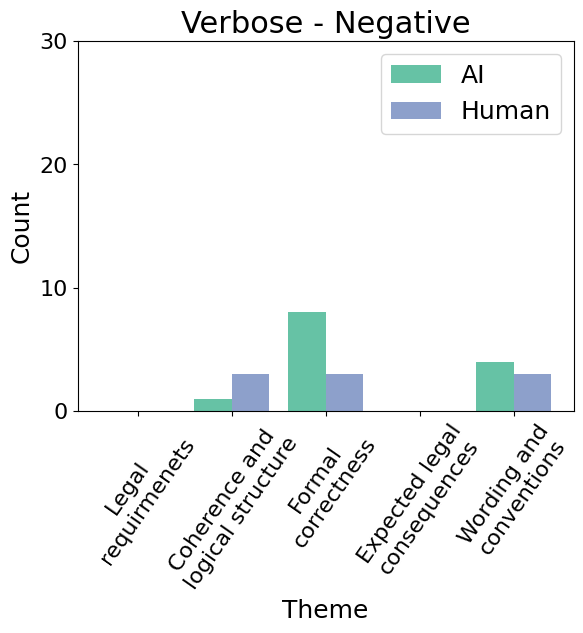}
\includegraphics[width=0.495\textwidth]{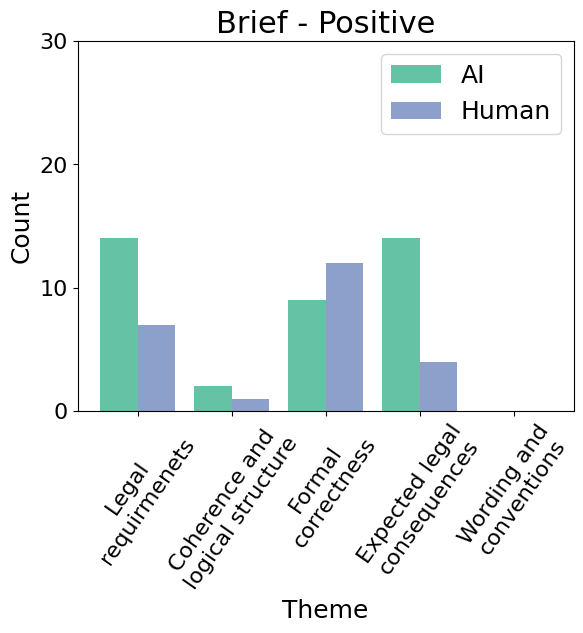}
\includegraphics[width=0.495\textwidth]{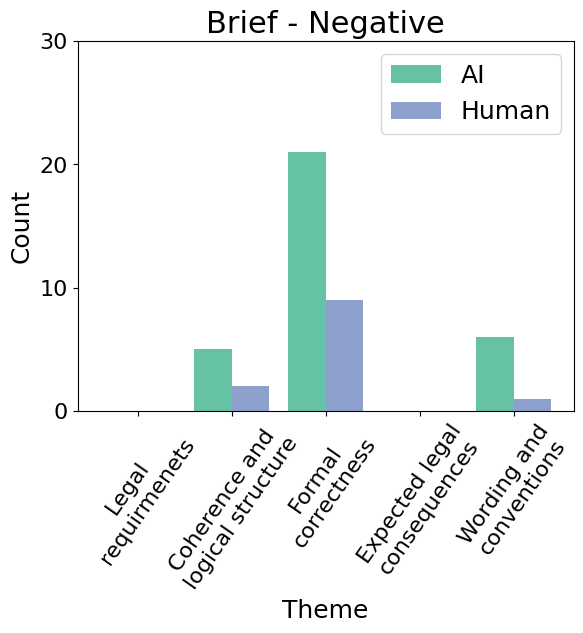}
\caption{Thematic analysis: Frequency of positive and negative mentions for each individual theme related to Correctness}\label{corr-themes}
\end{figure}

Figure \ref{corr-themes} shows that the AI-generated document has a lower ratio of positive mentions and a much higher ratio of negative ones. Note that the results of thematic analysis are not tested for statistical significance, but intended to demonstrate patterns and trends. The thematic analysis aligns with the outcome of the quantitative evaluation presented above.
Furthermore, there is a clear trend in Figure \ref{corr-themes} where the negative mentions of themes for documents labeled as AI-generated outnumber the negative mentions of the themes for documents labeled as human-crafted. Conversely, positive mentions are less common for documents when labeled as AI-generated. Interestingly, when the Brief document was labeled as generated by AI, there were numerous positive mentions of the \emph{legal requirements} and the \emph{expected legal consequences} themes. We found these themes often mentioned in conjunction with a negative comment. Although the evaluator had reservations about a document, in their opinion, it met the legal requirements and would have caused the expected legal consequences.


\subsection{Language Quality}
\label{language}
\begin{figure}
\centering
\includegraphics[width=0.495\textwidth]{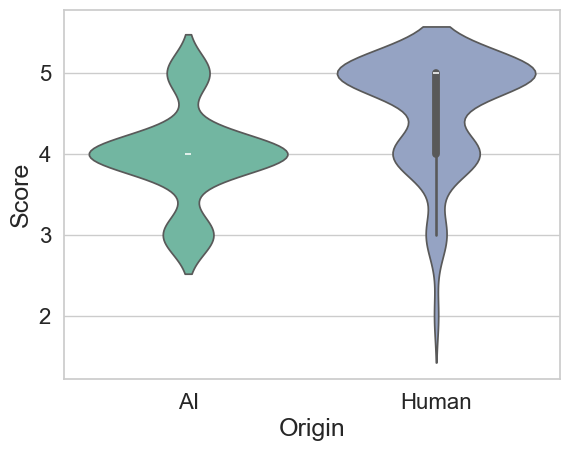}
\includegraphics[width=0.495\textwidth]{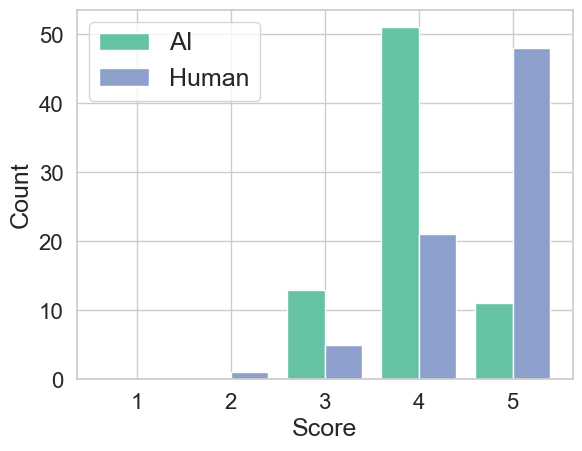}
\includegraphics[width=\textwidth]{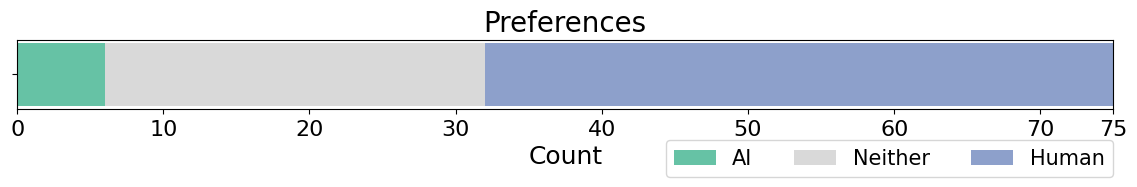}
\caption{The figure summarizes the participants' preferences between the documents when labeled as AI-generated versus human-crafted in terms of their Language Quality. The top two charts show the distribution of scores awarded to the documents carrying the ``AI`` (green) or ``human`` (blue) labels (1--worst; 5--best). The bottom chart presents the results of the side-by side comparison, showing how many times each of the labels was preferred (if any). Overall, a clear preference for the documents labeled as human-crafted over those labeled as AI-generated can be observed.}\label{lang-overall-dist}
\end{figure}

Figure \ref{lang-overall-dist} presents the overall evaluation of the documents in terms of their language quality, focusing on whether a document was labeled as human-crafted or AI-generated. The participants preferred a document when it was presented as human-crafted over a document labeled as AI-generated. Specifically, the mean evaluation of the two documents when labeled as human-crafted was 4.55 as compared to 3.97 when the same documents were marked as AI-generated. The side-by-side comparison presents a clear message when the human-crafted designation yielded 43 preferences over the mere 6 of the documents labeled as AI-generated (26 neither). This difference is statistically significant by Fisher exact test ($p<10^{-4}$ using Bonferroni correction for 4 comparisons).

\begin{figure}
\centering
\includegraphics[width=0.495\textwidth]{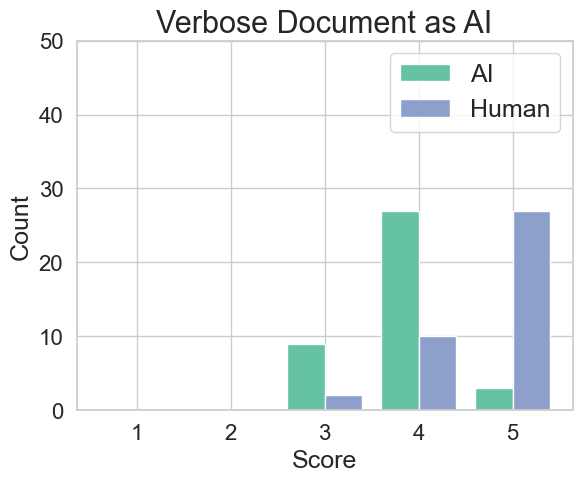}
\includegraphics[width=0.495\textwidth]{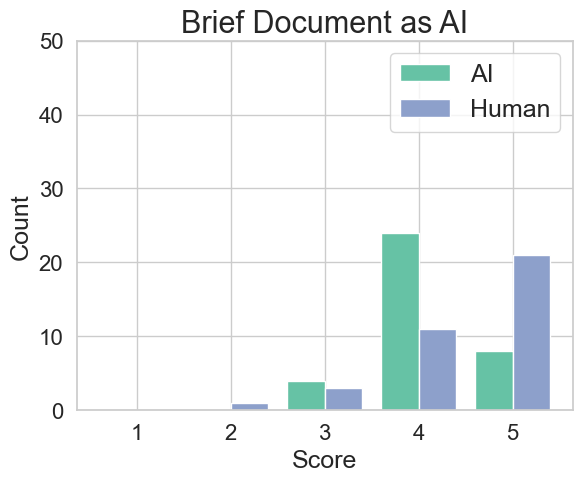}
\includegraphics[width=\textwidth]{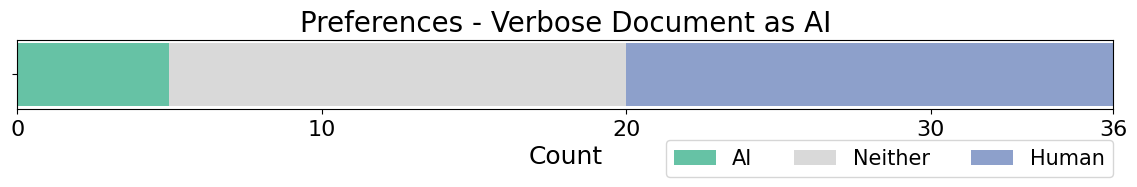}
\includegraphics[width=\textwidth]{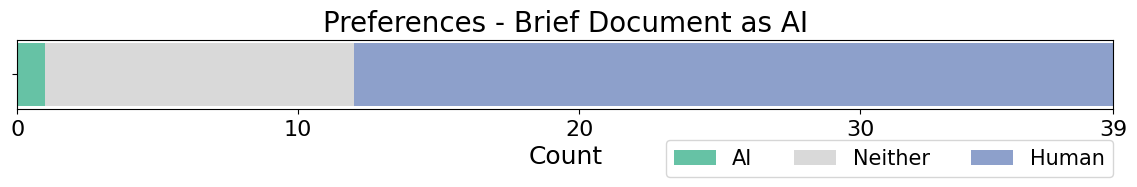}
\caption{The figure summarizes the participants' preferences between the documents when labeled as AI-generated versus human-crafted in terms of their Language Quality when the document is taken into account (Brief and Verbose). The top two charts show the distribution of scores awarded to the documents carrying the ``AI`` (green) or ``human`` (blue) labels (1--worst; 5--best). The bottom charts present the results of the side-by side comparisons, showing how many times each of the labels was preferred (if any) per document.}\label{lang-document-dist}
\end{figure}

Figure \ref{lang-document-dist} provides insight into the interaction between the AI/human-authored designation and the individual documents. Both documents appear to be preferred by the participants when labeled as human-crafted over AI-generated. On average, the Brief document was rated at 4.44 when labeled as coming from a human and at 4.11 when labeled as coming from an AI. For the Verbose document, the difference was even more pronounced (4.64 human versus 3.85 AI). In terms of the side-by-side comparisons, the Verbose document (labeled as AI-generated) was preferred only 1 time. In contrast, the Brief document marked as human-crafted was deemed of higher language quality 27 times (11 neither). When the Brief document was marked as produced by AI, it was preferred in 5 instances, while the Verbose document was preferred 16 times (i.e., 15 neither).

\begin{figure}
\centering
\includegraphics[width=0.495\textwidth]{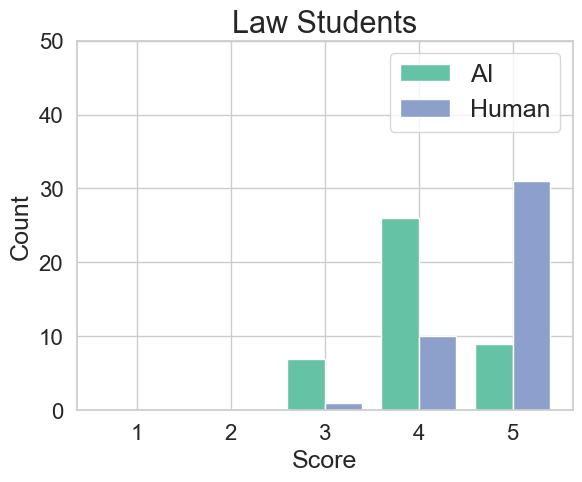}
\includegraphics[width=0.495\textwidth]{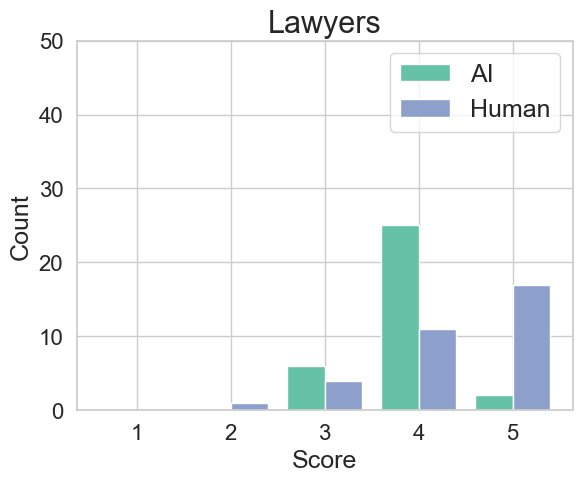}
\includegraphics[width=\textwidth]{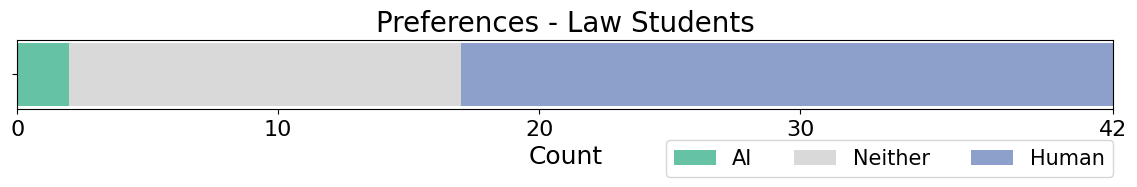}
\includegraphics[width=\textwidth]{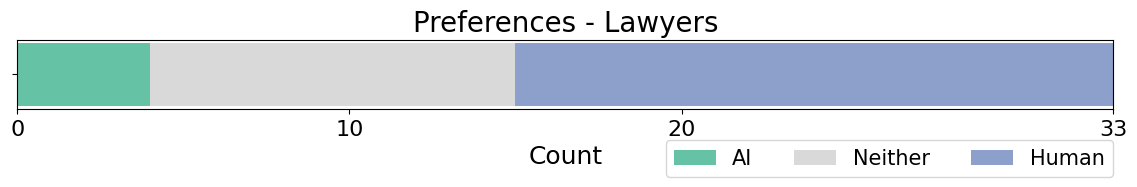}
\caption{The figure summarizes the participants' preferences between the documents when labeled as AI-generated versus human-crafted in terms of their Language Quality when the participants' background is taken into account (a law student and a lawyer). The top two charts show the distribution of scores awarded to the documents carrying the ``AI`` (green) or ``human`` (blue) labels (1--worst; 5--best). The bottom charts present the results of the side-by-side comparisons, showing how many times each of the labels was preferred (if any) per participants' background. It appears the AI-generated label has larger effect on the law students as compared to the lawyers.}\label{lang-background-dist}
\end{figure}

Figure \ref{lang-background-dist} focuses on the interaction between the participants' background (i.e., law student versus lawyer) and the AI/human designation of a document. Note that the result is not statistically significant by the Fisher exact test ($p=0.388$). The law students appeared to favor a document marked as human-crafted (4.71 labeled as human versus 4.05 for AI). Further, the law students preferred the document marked as coming from a human in 25 instances compared to only 2 preferences for a document presented as generated by AI (15 neither). As in the case of the correctness criterion, the lawyers appeared to have been less susceptible to the AI/human designation while still preferring a document when perceived as coming from a human (4.33 human vs 3.88 AI). They preferred a human-authored document in 18 cases compared to 4 instances when a document marked as AI-generated was preferred (11 neither).

To further understand the participants' preferences, we again conducted a thematic analysis of their explanations of the submitted document scores regarding their language qualities. The methodology is described in Subsection \ref{survey}. We identified four prominent themes in the explanations:

\begin{itemize}
    \item \emph{Grammar} -- comments on whether the whole document is grammatically correct or contains grammatical errors, such as the repeating reflexive pronoun in the Brief document.
    \item \emph{Stylistics} -- comments on using the correct professional terminology. 
    \item \emph{Structure} -- comments about whether the document contained all necessary structural parts (e.g., the header), whether the paragraphs were logically linked, whether the sentences were of adequate length and whether the document was formatted correctly. 
    \item \emph{Clarity} -- comments about the accuracy of expressions and overall comprehensibility of the document.
\end{itemize}

\noindent Each theme could be mentioned positively, e.g.,

\begin{quote}
    \begin{itemize}
        \item[] ``I found no grammatical errors.''
        \item[] ``Stylistically speaking, the document is fine.''
    \end{itemize}
\end{quote}

It could also be mentioned negatively, such as

\begin{quote}
    \begin{itemize}
        \item[] ``I find the sentences difficult to understand.''
        \item[] ``The header should appear above the first paragraph of the document.''
    \end{itemize}
\end{quote}

\textit{Grammar} is a theme containing exclusively comments on grammatical correctness or errors. In this respect, it is worth noting the Brief document intentionally contained a single grammatical error - the double reversible pronoun \textit{'se'}, specificity of the Czech language. Thus, in the case of a negative comment on \textit{grammar}, the comments exclusively pointed out this error in wording.

In other cases, the \textit{grammar} was commented on positively as being flawless, such as

\begin{quote}
    \begin{itemize}
        \item[] "I did not notice any grammatical or stylistic errors"
        \item[] "The document contains no grammatical or linguistic errors."
        \item[] "The spelling of the text is flawless."
        \item[] "there are no spelling errors in the text."
    \end{itemize}
\end{quote}

The theme of \textit{stylistics} includes references to terminology and, most often, to the professionalism of legal language. Positive comments include 

\begin{quote}
    \begin{itemize}
        \item[] "No colloquial expressions are used (on the contrary, professional legal language is used)."
        \item[] "The text is stylistically correct, and professional legal language is used."
        \item[] "Stylistically, the text is excellent, in terms of professionalism, the text is also fine."
    \end{itemize}
\end{quote}

Negative comments on the theme appeared as well, such as 

\begin{quote}
    \begin{itemize}
        \item[] "Stylistically, I would modify the document so that the line does not end with a conjunction. Some sentences repeat words that could be replaced by alternatives."
        \item[] "Stylistically, there are some distracting elements, such as the excessive use of parentheses or repetition of words."
        \item[] "I consider it stylistically superfluous to use the demonstrative pronoun 'of this Agreement' if the Agreement is previously defined as an abbreviation."
    \end{itemize}
\end{quote}

The document \textit{structure} theme includes comments on missing structural parts of documents (such as headers), logical continuity of individual parts, sentence length and flow of text, document formatting and clarity (e.g. indentation). In a positive sense, the theme of \textit{structure} appears much more frequently in the Verbose document than in the Brief one. Specifically, these comments include,

\begin{quote}
    \begin{itemize}
        \item[] "Sentences were not unnecessarily long, and the complexity of the sentences did not exceed that of a standard legal text in difficulty."
        \item[] "The structure of the text is clear."
        \item[] "The structure and formatting further facilitate understanding of the text."
    \end{itemize}
\end{quote}

Negatively, the theme \textit{structure} is mentioned, for example, as

\begin{quote}
    \begin{itemize}
        \item[] "The sequence of the text, where only the 3rd paragraph mentions the debtor acknowledging the debt, seems illogical to me."
        \item[] "There is a lack of a header for better orientation as regards the parties of the acknowledgement in the text. Some sentences are too long."
        \item[] "For the sake of clarity, I would have placed the header ahead of the document, not in the text."
        \item[] "The way the text is indented is slightly distracting to me."
    \end{itemize}
\end{quote}

The theme of \textit{clarity} is more often mentioned positively for a document labeled as human-crafted, while AI-generated is more often mentioned negatively. In general, clarity comments refer to the accuracy and comprehensibility of the expression. In negative terms, the comments relate to the excessive complexity of the text and the possibility of making it simple through editorial intervention.



\begin{figure}
\centering
\includegraphics[width=0.495\textwidth]{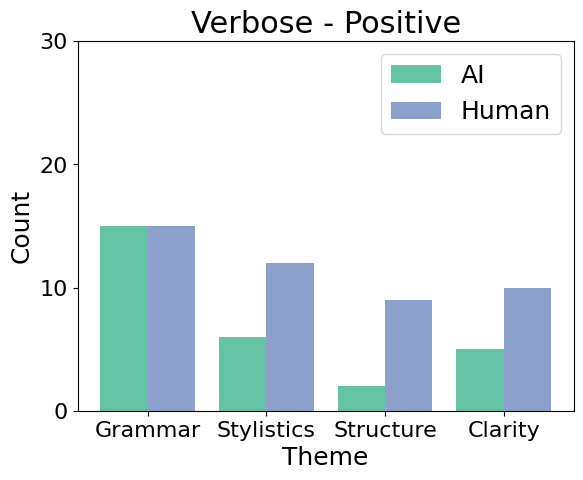}
\includegraphics[width=0.495\textwidth]{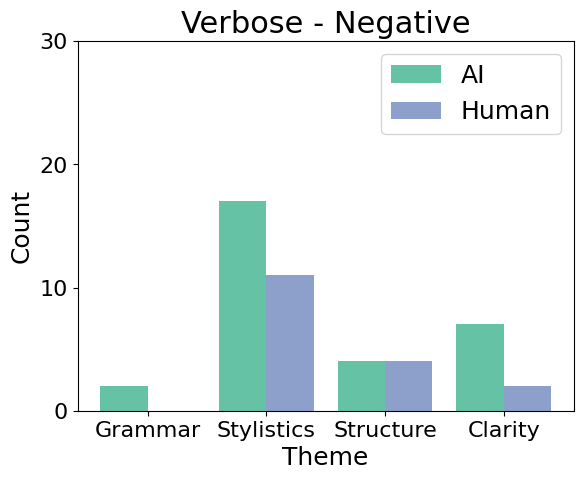}
\includegraphics[width=0.495\textwidth]{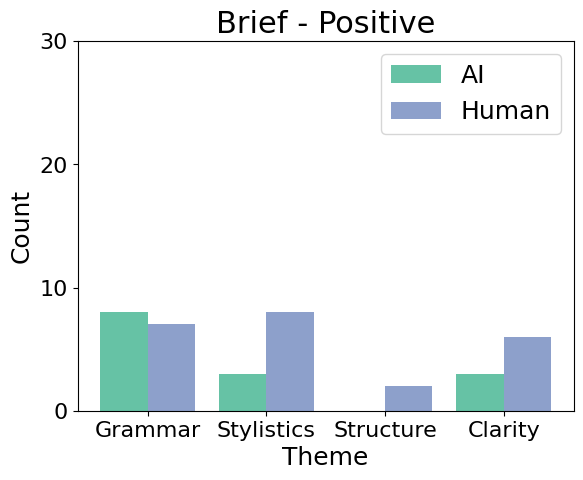}
\includegraphics[width=0.495\textwidth]{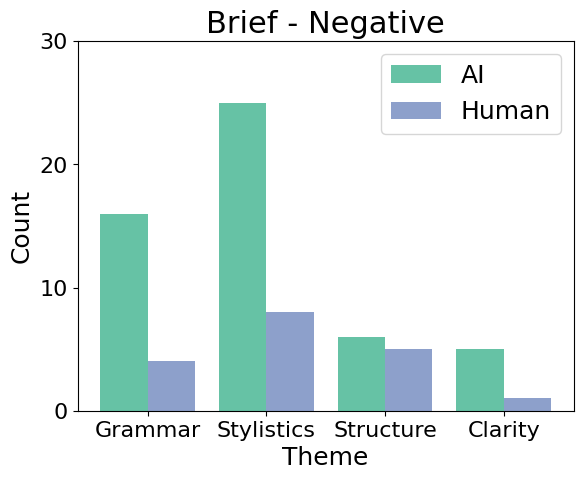}
\caption{Thematic analysis: Frequency of positive and negative mentions for each individual theme related to Language Quality,}\label{lan-themes}
\end{figure}

 The ratio of positive versus negative mentions confirms the results of the quantitative analysis. Note that the results of thematic analysis are not tested for statistical significance, but intended to demonstrate patterns and trends. The detailed results of the thematic analysis are shown in Figure \ref{lan-themes}. There, we can see the high prevalence of negative mentions for the documents when labeled as AI-generated (both Verbose and Brief) across all the themes. Once the documents are labeled as human-crafted, positive mentions prevail. We also find a notable difference in the ratings for the Brief document. When labeled as AI-generated, more negative comments on language quality in each category appeared. 


\subsection{Possibility of Automated Generation}
\label{automation}
After evaluating the documents, the participants were asked to answer an open-ended question about the possibility of a fully automated generation of documents similar to those they had just seen. The overwhelming majority, 93\%, answered that they believed this would be possible. Only 5\% remained uncertain, and merely 2\% believed this to be impossible.

We further explored the reasons that led the participants to believe in the possibility of future automation. Responses to this question were thematically coded following the methodology described in Subsections \ref{survey}. The codes were then collated into five higher-level themes as follows:

\begin{itemize}
    \item \emph{Documents Indistinguishable} -- the evaluators claimed that they found human-crafted and AI-generated documents practically indistinguishable, and thus full automation was possible.
    \item \emph{Expected Legal Consequences} -- the evaluators argued that if, despite some reservations, the document can produce the expected legal consequences, then it is very likely that full automation of such documents is possible.
    \item \emph{Similarity to Forms} -- the evaluators argued that it is already possible to mechanize some legal actions by using forms (e.g. simple contracts), and using LLMs is akin to more technologically advanced forms.
    \item \emph{Sufficient Content} -- the evaluators commented that they found the content of the document sufficient and, as a result, did not see a problem with full automation.
    \item \emph{Lower Cost} -- the evaluators commented on the cost/benefit by comparing the time and money saved by automatic generation, even at the cost of the document not being perfect.
\end{itemize}

\begin{figure}
\centering
\includegraphics[width=.7\textwidth]{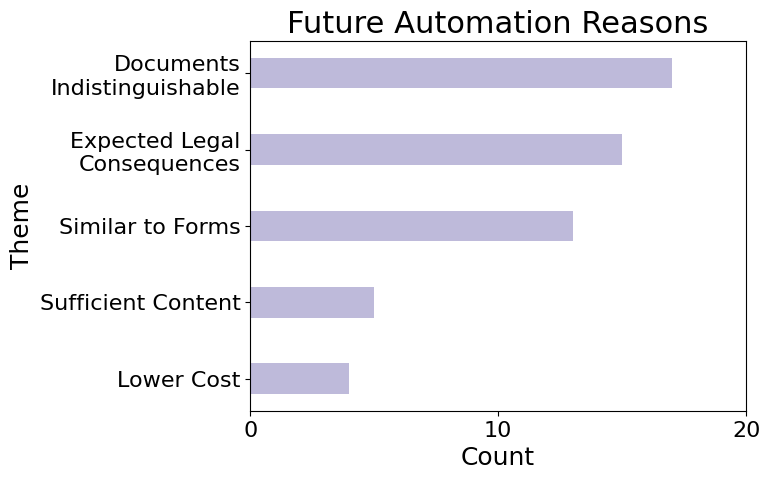}
\caption{Thematic analysis: Frequency of reasons for possibility of future automation}\label{automation-poss}
\end{figure}

\noindent As can be observed from Figure \ref{automation-poss}, the most common reasons were the \textit{indistinguishability} of the two documents (human-crafted and AI-generated), then the fact that both documents were capable of producing the \textit{expected legal consequences}, and then the argument stating that this is just a natural evolution of \textit{form-based} document automation.

A general conclusion in the context of the negative evaluation of both the language and the correctness of the document, which was labeled AI-generated, is that even though the participants evaluated such a document more negatively, they believe it would still hold up in practice. Therefore, they consider the automation of legal documents to be very realistic. Despite the evaluators' reservations about the linguistic and legal quality of a document labeled as AI-generated, it can be assumed from their answers that the document is, in their opinion, sound and can be fully automated.

\section{Discussion}
\label{discussion}
The results of our study reveal several significant trends that are of utmost importance to the legal profession and researchers in AI and law.
 
Before we approach individual issues, we must address the overarching trend, which, while not the subject of our study, is noteworthy. The documents were prepared to be of comparable quality, terminology, and structure, established via cross-validation within the authors' team and by approaching an experienced attorney outside the authors' team. However, the Verbose document, on average, outperformed the Brief document in the evaluation, both in terms of correctness and language quality. A detailed investigation of the trend reveals that the Brief document labeled as human-crafted outperforms the Verbose document labeled as AI-generated. However, the difference is not significant enough to overcome lawyers' and law students' general preference for the Verbose document. Such a preference may be rooted in the tendency of legal professionals to associate length with precision.

Additionally, it reflects the implicit nature of legal knowledge. Civil law, which the acknowledgement of debt we used in the study is the prime example, uses dispositive norms. Unlike cogent norms, dispositive norms allow contracting parties to reach an agreement which is, in detail, different from what is stipulated in the respective Act. It might be that lawyers preferred the Verbose document in their expectation that it captures the will of the parties expressly and fully and does not leave any part of the contract tacitly bound to the Civil Code. While we noted the general preference for the Verbose document over the Brief one, the AI-generated Verbose document was rated worse than the Brief document labeled human-crafted. Lawyers expect longer human-crafted documents to be a sign of eloquence, while longer AI-generated documents indicate uncertainty and an inability to operate in the legal domain. The preference for Verbose documents should be investigated further regarding underlying motivations.

In terms of individual trends within the scope of this study, participants were more critical when scoring and commenting on language quality over correctness. Negative comments about language were more prevalent for both Brief and Verbose documents when presented as AI-generated. The risks of LLMs, associated mainly with the prevalent hallucinations, are well-known among both lawyers and law students. As such, participants may have entered the survey expecting to see these issues manifest when dealing with AI-generated documents. Since both documents were prepared to meet legal requirements, and the acknowledgement of debt is straightforward, participants may have focused on language-related issues instead of correctness. Specialized legal jargon is not known for its brevity and conciseness. As a result, most participants can point out weaknesses based on individual stylistic preferences. \citet{Castelo2019} identified the aversion to algorithms as higher in tasks perceived as subjective.

The correctness is objective, while language issues are affected by individual preferences, such as using specific expressions. Such an aversion to distinguishing between specific and objective tasks may be another explanation for negative comments directed at the documents' language quality when presented as AI-generated. Our interpretation is further supported by the heightened occurrence of negative comments tied to more subjective aspects of correctness. For example, the 'formal correctness' of the documents presented as AI-generated drew a lot of negative comments, as it is again a subjective category where individual preferences matter.

In support of this interpretation, we noted that none of the documents, either Brief or Verbose, whether labeled as AI-generated or human-crafted, received a negative comment regarding legal requirements and expected legal consequences. While individual preferences may influence other properties, these are the most objective, as they are directly based on the law. Lawyers may object to subjective properties, but it is difficult to dispute the ability of documents to give rise to specific legal consequences. The interpretation may serve as a stepping stone for other studies with a more nuanced approach towards evaluating documents' properties. We focused on correctness and language quality, but both broadly defined categories contain subjective sub-categories. Mapping the impact of these variables would have a profound impact on the further use of LLMs within the legal domain.

Documents labeled as human-crafted were consistently rated better than AI-generated documents in terms of correctness. However, there is nuance to these results, which warrants further attention. Thematic analysis has shown lawyers criticizing the AI-generated Brief document in terms of its coherence, logical structure, and formal correctness. The prevalence of negative comments in these areas was lower in the AI-generated Verbose document. Inherent negative expectations regarding the language capabilities of LLMs may cause this. The legal jargon often contains complicated sentences with conditions and caveats. The lawyers expect AI to be unable to use the language proficiently enough to structure a legal text and perceive the AI-generated Brief document as supportive evidence for their stance.

On the other hand, quite surprisingly, the Brief document labeled as AI-generated attracted more positive comments on legal requirements and expected legal consequences than its Brief counterpart labeled human-crafted. Well aware of risks related to hallucinations of factual information, lawyers were probably surprised by the factuality of the Brief Document presented as AI-generated. Lawyers are prepared to encounter imperfect or faulty AI-generated documents. As a result, they felt the urge to comment positively on what they otherwise perceived as 'normal' or 'expected' within human-crafted documents. The aspect of expectation was presented even in the human-crafted Verbose document, receiving more negative comments regarding coherence and logical structure than the AI-generated Verbose document. Lawyers seem to have low expectations of AI-generated documents. On the other hand, their opinion of human-crafted documents may be affected by the projection of how the document would have looked should they have drafted it. Such a projection can carry with it a negative sentiment.

Additionally, documents labeled as human-crafted received better overall scores and more positive and less negative comments. These findings are in line with the existing literature on algorithmic aversion. While the previous research reported lower evaluation of AI-generated content in case of Airbnb profiles \citep{Jakesch2019}, emails \citep{Liu2022}, artworks \citep{Ragot2020}, music \citep{Shank2023}, translations \citep{Asscher2023}, news articles \citep{Waddell2018} or health prevention messages \citep{Lim2024}, we identified the lower evaluation of AI-generated legal content. Automated legal content generation may also lack the required level of personalized care about one's legal issues. The outcome aligns with concerns in the healthcare sector as \citet{Longoni2019} mention lack of perceived personalization as one source of resistance to the deployment of artificial intelligence.

Although documents labeled as AI-generated were consistently rated worse, and the thematic analysis yielded more negative comments on both their linguistic and legal quality, the results of the open-ended question about possible future automation reveal that practitioners certainly envisage full automation of (at least) simple legal documents such as the acknowledgement of debt. While there is a noticeable scepticism towards AI-generated legal documents, lawyers consider full automation in the future almost certain. The interpretation holds regardless of the lower rating of documents labeled as AI-generated in both correctness and language quality.

\section{Implications for Legal Practice}
\label{implications}
The disruptive potential of LLMs is immense. The most often appearing opportunities are decreasing costs associated with legal services and increasing access to justice. ChatGPT's appeal partly lies in the democratization of legal services. Lawyers' services are often perceived as prohibitively costly. Laypeople already use LLMs to address their legal needs \citep{Hagan2024}, and the trend is likely to grow.

Two major issues stand in the way: 
\begin{enumerate}
    \item the accuracy of the provided information and generated documents, and
    \item perceptions and attitudes towards AI-generated content.
\end{enumerate} 

These two are communicating vessels. Inaccuracies, either real or perceived, can hamper the use of LLMs in the legal domain. Our study shows that the perception of AI-generated documents must be studied separately from their objective performance. The performance of LLMs in individual tasks, either low-level or complex, is just one part of the story. Even objectively accurate documents can be perceived negatively when their recipients believe they are AI-generated.

With legal services prohibitively expensive, lower-income groups are likely to be excluded from reaching human legal assistance. Such people may often turn to legal aid, where LLMs will be used to address the known bottlenecks, or to self-help with LLMs augmenting the knowledge. In both instances, the underperformance of LLMs can hamper their chances of success.

However, even if appropriately used, the perception of lower-quality of AI-generated content also harms their chances. Additionally, people from lower-income groups may not have to approach government agencies or courts with AI-generated content to be negatively affected. The mere perception that a person is using AI-generated content can lower their chances of addressing their legal needs. In a broader context, unequal viewing of similar legal documents with different origins (AI, human) may, in turn, contribute to increased social inequalities.

Our results have shown that the label of AI-generated content carries negative associations, such as broad indication of the document's lower quality or general inadequacy to cause legal consequences.

As a result, mandatory disclosure of using LLMs may manifest as a conflict between fairness and transparency. Algorithmic aversion may become another bias, which will require practical measures to address its detrimental impact. The high-level requirements for fair and just trial may not be enough to address implicit biases, whether racial, gender, or AI-related. Instead of better access to justice aided by ChatGPT, LLMs, and AI in general, we may be left with implicit constraints placed upon those needing legal assistance the most.

\section{Limitations}
\label{limitations}
While our study provides valuable insight into preferences and perceptions regarding AI-generated legal documents, limitations must be acknowledged.

The participants of our study, law students and lawyers, are a relevant demographic for evaluating legal documents. However, it is important to note that the results may not generalize well in other contexts, such as among officials, judges, prosecutors, or the general population. These groups may perceive AI-generated documents differently due to their varying levels of prior knowledge and different encounters with legal issues.

Our study's limited number of participants (n=75) may have introduced bias to the results. As a result, some of the reported results are not statistically significant. Furthermore, the evaluation of the correctness and language quality does not sufficiently capture the various properties of the documents. A more comprehensive analysis is necessary to address this limitation.

Our study evaluated the acknowledgement of debt. While the document is commonly used and is a relevant case study, our findings may not extend to other legal documents. Additionally, acknowledgement of debt is simple. Different document types have more complex legal requirements or linguistic and structural properties. Exploring a more comprehensive range of legal documents would lead to more robust data.

We did not address the issue of prior use or exposure to AI when recruiting the participants. Such a variable may prove significant regarding expectations and perceptions of AI-generated documents.

Finally, we evaluated the documents in the Czech language. As a demographic, Czech lawyers and law students may have been less exposed to LLMs and less accustomed to their use than other nationalities. Additionally, \citet{Manvi2024} reported LLMs to be geographically biased because of used training corpora and demonstrated language capabilities. Expectations tied to AI-generated documents in Czech - as one of the minor languages - may have influenced the results reported above.

Addressing these limitations in future research will be crucial in achieving more generalizable and potentially practical results.

\section{Conclusions and Future Work}
\label{conclusion}
Our study shed light on how lawyers perceive legal documents labeled as AI-generated compared to documents labeled as human-crafted. Participants rated documents labeled as AI-generated documents worse than their counterparts presented as human-crafted.

Participants were more critical of language quality than correctness, with AI-generated documents receiving more negative comments. The aversion to AI-generated legal content aligns with existing research on algorithmic aversion in other domains. Documents labeled as human-crafted received higher overall scores and more positive comments, reflecting a possible general preference for the involvement of human experts in producing legal content.

Despite the prevalent negative perception, our study reveals significant optimism among the participants about the future of automated legal content production. They highlighted factors such as the indistinguishability between human-crafted and AI-generated content and the ability of AI-generated content to produce legal consequences, indicating a potential shift in perception in the future.

Our study is the first effort to understand the perception of AI-generated content on its recipients in the legal domain. Our study confirms the negative perception of AI-generated content and the general preference for human-crafted documents. 

While our findings are significant, they should be a starting point for further research. The potential repercussions of providing legal services—particularly for lower-income groups—and the mandatory labeling of AI-generated documents require careful evaluation. Further research must focus on other populations, larger samples, and more complex documents, as well as a nuanced understanding of documents' objective and subjective properties and the prior knowledge of intended recipients.

The perception of AI-generated documents still needs to be studied regarding the variables and their subsequent impact on legal practice.

\bmhead{Acknowledgements}
We thank Bettina Bacher, and the two anonymous reviewers, for their helpful comments on previous versions of this paper.
Work of Jakub Harasta and Tereza Novotná was supported by the Grant Agency of Masaryk University (GAMU) project “Forensic Support for Building Trust in Smart Software Ecosystems” (no. MUNI/G/1142/2022).

\bibliography{sn-bibliography}

\end{document}